\documentclass[twocolumn,prd,nofootinbib,longbibliography]{revtex4-1}
\pdfoutput=1
\usepackage{amsmath}
\usepackage{graphicx}
\newcommand{\beq}{\begin{equation}}
\newcommand{\eeq}{\end{equation}}

\def\aap{Astron.\ Astrophys.}
\def\apjs{Astrophys.\ J.\ Supp.}
\def\physrep{Phys.\ Rep.}
\def\nar{New Astron.\ Rev.}
\def\mnras{M.N.R.A.S.}

\begin{document}

\title{Pure E and B polarization maps via Wiener filtering}
\author{Emory F. Bunn}
\affiliation{Physics Department, University of Richmond, Richmond, VA  23173, USA}
\author{Benjamin Wandelt}
\affiliation{Sorbonne Universit\'{e}, UPMC Univ Paris 6 et CNRS, UMR 7095, Institut d'Astrophysique de Paris, 98 bis bd Arago, 75014 Paris, France\\
Sorbonne Universit\'{e}, Institut Lagrange de Paris (ILP), 98 bis Boulevard Arago, 75014 Paris, France\\
Departments of Physics and Astronomy, University of Illinois at
Urbana-Champaign, Urbana, IL 61801, USA}

\begin{abstract}
In order to draw scientific conclusions from observations of cosmic microwave background (CMB) polarization, it is necessary to separate the contributions of the E and B components of the data. For data with incomplete sky coverage, there are ambiguous modes, which can be sourced by either E or B signals. Techniques exist for producing ``pure'' E and B maps, which are guaranteed to be free of cross-contamination, although the standard method, which involves constructing
an eigenbasis, has a high computational cost. 
We show that such pure maps can be thought of as resulting from the application of a Wiener filter to the data. 
This perspective  leads to far more efficient methods of producing pure maps. 
Moreover, by expressing the idea of purification in the general framework of
Wiener filtering (i.e., maximization of a posterior probability), it
leads to a variety of generalizations of the notion of pure E and B maps, e.g.,  
accounting for noise or other contaminants in the data as well as correlations with temperature anisotropy.
\end{abstract}

\maketitle

\section{Introduction}

Characterization of the polarization of the cosmic microwave background (CMB) radiation is an extremely high priority in cosmology. Polarization has been measured by the Planck \cite{planck2015maps,planck2015cl} and
WMAP \cite{wmap9yrresults} satellites, as well as by a number of suborbital telescopes (e.g., \cite{dasi,cbi,polarbear,quiet}).

Polarization maps can constrain cosmological models in a variety of ways, perhaps the most exciting of which is the potential detection of a stochastic background of gravitational waves produced during an inflationary epoch (e.g., \cite{guzzetti}). Detection of this signal would constitute a direct confirmation of inflation and a measurement of the inflationary energy scale, with revolutionary effects on the field.

Detection of this signal, as well as other phenomena such as gravitational lensing, in CMB polarization maps depends on the fact that a polarization map can be represented as the sum of a scalar E component and a pseudoscalar B component (the spin-2 analogues of curl-free and divergence-free vector fields respectively) \cite{kkslett,selzal,zalsel,kks}. To linear order in perturbation theory, scalar density perturbations produce only the E component, leaving the B channel clear as a probe of other phenomena such as inflationary gravitational waves.
The E-B decomposition is also important in the analysis of surveys
of weak gravitational lensing (e.g., \cite{weaklensingreview} and
references therein).

Because the B component is predicted to be at least an order of magnitude smaller than the E component over all angular scales, detection of B modes requires clean separation of the two components. For data with complete sky coverage, leakage-free separation of the two components is straightforward. However, in a data set with partial sky coverage there are ``ambiguous modes'' that cannot be uniquely assigned to either E or B \cite{bunndetect,lct,lewis,BZTO,bunnpurify}. 

Although the E-B decomposition is not unique for data with incomplete sky coverage, one can uniquely decompose any polarization map into three components, dubbed \textit{pure E}, \textit{pure B}, and \textit{ambiguous}
\cite{lct,BZTO}. The pure E (B) component lies in the orthogonal complement of the vector space of all B (E) modes. The ambiguous component is orthogonal to both the pure E and pure B spaces. This decomposition has the advantage that any signal found in the pure B map is guaranteed to have come from actual B modes.

Original work on the pure-ambiguous decomposition involved the construction of orthonormal bases for the various subspaces, which was a slow and cumbersome procedure. An alternative view emerges when considering the CMB polarization analysis problem in the context of  Gibbs sampling \cite{WandeltGibbs,JewellLevinAnderson,LarsonGibbs}. Since Gibbs sampling infers the posterior statistics of the all-sky signal given the data, separating E and B modes is trivial for every sample. The resulting set of T, E, and B map samples represents the information the data contain. Gibbs sampling does not require mode decompositions but relies purely on optimal filtering of the data augmented to cover the entire sky. It is therefore natural to ask whether the E/B mode separation problem can be approached from the filtering point of view.

In this paper, we show that  the decomposition can be thought of as an application of  the familiar Wiener filter, which allows for a much faster implementation. In particular, because the filter can be expressed in terms of operations that are diagonal in either pixel space or spherical harmonic space, efficient techniques such as conjugate gradient solution and the messenger method \cite{messenger} can be applied.

In addition to providing an efficient method of computation, the Wiener filter approach places the pure-ambiguous decomposition into a context that allows it to be generalized in a variety of ways. This approach can be used to generate the original pure and ambiguous maps, but it can also be used to generate more useful generalizations of them that treat noise and other contaminants and the mask together 
in a unified way, with the mask regarded simply as a region of infinite noise. The resulting filtered pure maps thus suppress noise-dominated modes while simultaneously accounting for the mask.

The method also has a natural generalization to include correlations between temperature and polarization. In particular, it provides a natural way of ``purifying'' the E map from the portion that is correlated with temperature, giving a clear view of the new information contained in such a map.

There are many other methods of dealing with the problem of EB leakage.
Some work at the power spectrum level, without producing real-space
maps of the E and B modes \cite{smithpseudocl,smith2,smithzal,grain,challchon}. Others produce maps in real space of estimates of the derivatives of the 
purified polarization maps \cite{kimnaselsky,kim,zhao,bowyer}.
While these are potentially quite valuable, they differ from the approach taken herein, which aims to produce filtered maps of the actual polarization measurements, rather than their derivatives.
Ref. \cite{ferte} assesses the merits of these methods for power-spectrum
estimation.
Wavelet-based methods \cite{caofang,leistedt,rogers} have also been developed.
These approaches require a certain amount of tuning (e.g., careful choice of scale-dependent masks),
whereas the Wiener approach allows all filtering to be performed
in a principled way from the data and a theoretical model.

The remainder of this paper is organized as follows. Section \ref{sec:setup} establishes notation and provides a brief review of some aspects of the EB decomposition. Section \ref{sec:noisefree} shows how the pure-ambiguous decomposition can be implemented as a quadratic minimization problem involving extension of the data into the masked region, allowing us to take advantage of efficient all-sky E-B separation. In section \ref{sec:wf}, we show that this approach generalizes in a natural way to the application of a Wiener filter. Section \ref{sec:tcorr} generalizes the previous results to include temperature-correlation. Section \ref{sec:examples} shows some examples of the method, and section \ref{sec:conclusions} contains some brief concluding remarks.

\section{Basic setup and notation}
\label{sec:setup}

Our data set consists of measurements of the linear polarization Stokes parameters $Q,U$ at a set of $N_{\rm obs}$ pixels covering part of the sky. We will consider the correlation with temperature (Stokes $I$) measurements in Section \ref{sec:tcorr}.
As usual, each measurement contains both signal and noise:
\beq
d_j = s_j + n_j.
\eeq
We assume Gaussian noise with covariance matrix $\mathbf{N}$.
In addition to instrumental noise, $n_j$ can include the effects of
residual foregrounds or other contaminants in the data.

The index $j$ labels both the pixel location and the Stokes parameter,
so the vector $\vec d$ has dimension $2N_{\rm obs}$.
The signal can be expressed as a spherical harmonic expansion
\beq
s_j=s_j^E+s_j^B=\sum_{l,m}(a_{E,lm}Y_{E,lm}^Z(\hat r_j) +a_{B,lm}Y_{B,lm}^Z(\hat r_j)).
\eeq
Here $\hat r_j$ is the location of the pixel corresponding to
measurement $j$ and $Z\in \{Q,U\}$ is the Stokes parameter of that measurement.
The functions $Y$ are related to the spin-2 spherical harmonics:
\begin{align}
Y_{E,lm}^Q&=Y_{B,lm}^U=-{1\over 2}({}_2Y_{lm}+{}_{-2}Y_{lm}),\\
Y_{B,lm}^Q&=-Y_{E,lm}^U = -{1\over 2}({}_2Y_{lm}-{}_{-2}Y_{lm}).
\end{align}
If the data cover a small enough region that the flat-sky approximation is appropriate, then the spherical harmonics can be replaced with plane waves and fast Fourier transforms may be used.
See \cite{BZTO} and references therein for further details.

The signal vector can thus be written
\beq
\vec s = \mathbf{Y}_E\vec e + \mathbf{Y}_B\vec b,
\eeq
where the vectors $\vec e$ and $\vec b$ contain the coefficients 
$a_{E,lm}$ and $a_{B,lm}$ respectively, and the matrices $\mathbf{Y}_Z$ contain the spherical harmonics evaluated at the pixel locations.
[To be specific, for each element $(\mathbf{Y}_Z)_{j\alpha}$,
$j$ labels the pixel location and Stokes parameter, and $\alpha$ labels
the index pair $(l,m)$.]

If we make the usual assumption that the data are derived from a 
statistically isotropic, parity-respecting Gaussian random process, then the theory is completely described by the signal covariance matrix 
\beq
\mathbf{S}\equiv \langle \vec s\vec s^\dag\rangle
=\langle \vec s^E\vec s^{E\dag}\rangle+\langle \vec s^B\vec s^{B\dag}\rangle
\equiv\mathbf{S}_E+\mathbf{S}_B,
\eeq
where
\beq
\mathbf{S}_Z
=\mathbf{Y}_Z\mathbf{C}_Z\mathbf{Y}_Z^\dag.
\eeq
Here $Z\in \{E,B\}$, and the diagonal matrices $\mathbf{C}_{ZZ}$ contain the power spectra. 
Specifically, $(\mathbf{C}_{ZZ})_{\alpha\beta}=\delta_{\alpha\beta}C_l^{ZZ}$, where
$l$ is the multipole index corresponding to $\alpha$ and $C_l^{ZZ}\equiv
\langle a^*_{Z,lm}a_{Z,lm}\rangle$ is the usual power spectrum.

If the data vector $\vec d$ covers the whole sky, then the matrices
$\mathbf{Y}_E,\mathbf{Y}_B$ span orthogonal subspaces, and hence the 
$E$ and $B$ signals can be estimated independently of one another.
In maps with incomplete sky coverage, however, this is not the case: there
are ``ambiguous modes'' that lie in both subspaces simultaneously.
It is impossible to say whether power in such a mode came from E or B modes. However, we can divide the signal vector space into \textit{three}
orthogonal subspaces, dubbed pure E, pure B, and ambiguous spaces.
The pure E space is the orthogonal complement of the space spanned by
$\mathbf{Y}_B$: that is, it consists of modes that are orthogonal (over 
the observed region) to all possible B maps. The pure E space is 
defined similarly, and the ambiguous space is orthogonal to both of these.
If the data set $\vec d$ is projected onto the pure B subspace, then all signal from E modes will be mapped to zero -- that is, any power seen in this pure B map (beyond the noise) is guaranteed to have come from B modes.

%% Because this definition centers on the notion of orthogonality, it
%% depends on the choice of an inner product.
%% In \cite{BZTO}, the standard inner product $(\vec d,\vec d')=\vec d\cdot
%% \vec d' = \sum_j d_jd'_j$ was used. Note, however, that this is not the only choice: the division into pure and ambiguous modes can be made with any inner product. This observation will become important when we examine the connection between the pure-ambiguous decomposition and the Wiener filter.

\section{Fast purification}
\label{sec:noisefree}

Early work focused on finding bases for the pure and ambiguous spaces by solving an eigenvalue problem, and this method was implemented in the analysis of the BICEP2 data \cite{bicep2}. However,
this procedure is slow and cumbersome. In this section we show an efficient way of finding the pure maps corresponding to a given data set.

Let us ignore noise for the moment and consider just the effects of incomplete sky coverage. 
Our data contains observations at $N_{\rm obs}$ points on the sky. Let us embed the $2N_{\rm obs}$-dimensional data vector into a larger $2N_{\rm pix}$-dimensional vector space whose pixels cover the entire sky.
Let $\mathbf{M}$ be the ``mask operator'' which orthogonally projects onto the space of observed pixels: 
\beq
\mathbf{M}_{jk}=\begin{cases}\delta_{jk} & 
\mbox{if $j$ corresponds to an observed pixel}\\
0 & \mbox{otherwise.}
\end{cases}
\eeq
We set the data vector to zero for all unobserved pixels,
so $\mathbf{M}\vec d=\vec d$. 

One way to produce the pure B map corresponding to $\vec d$ is to find the extension of $\vec d$ into the unobserved region that minimizes the total B power. We will now describe this procedure in detail and prove that it is equivalent to the usual definition of the pure B map. Naturally, an equivalent statement applies to the pure E map.

Let $\vec\delta$ be an extension of $\vec d$ into the observed region, so that 
\beq
\mathbf{M}\vec\delta=\mathbf{M}\vec d.
\label{eq:constraint}
\eeq
 Because $\vec\delta$ is an all-sky map, we can unambiguously decompose it into E and B components by applying projection operators $\mathbf{P}_E$ and $\mathbf{P}_B$. 
We choose $\vec\delta$ to minimize the quantity
\beq
\phi=(\mathbf{P}_B\vec\delta)^2
\eeq
subject to the constraint (\ref{eq:constraint}). The result is
\beq
\mathbf{P_B}\vec\delta = \mathbf{M}\vec\lambda,
\label{eq:lagrangemultiplier}
\eeq
where $\vec\lambda$ is a vector of Lagrange multipliers.
In other words, $\mathbf{P}_B\vec\delta$ lies entirely in the observed
(unmasked) part of the sky. Because $\mathbf{M}^2=\mathbf{M}$,
\beq
\mathbf{M}\mathbf{P}_B\vec\delta=\mathbf{P}_B\vec\delta.
\label{eq:mpd=pd}
\eeq

We know that
\beq
\vec\delta = \mathbf{P}_E\vec\delta+\mathbf{P}_B\vec\delta.
\eeq
We wish to show that the first term in this expression contains the E modes (pure and ambiguous), and the second contains the pure B modes. The two
terms lie in the E and B subspaces respectively, so all we have to show is that the second term is ``pure'' -- i.e. that it is orthogonal to all possible E modes over the observed region. Let $\vec\epsilon$ 
be an arbitrary E mode. The dot product of $\vec\epsilon$ with
$\mathbf{P}_B\vec\delta$, taken only over the observed pixels, is
\beq
\vec\epsilon^{\,\dag} \mathbf{M}\mathbf{P}_B\vec\delta=\vec\epsilon^{\,\dag}
\mathbf{P}_B\vec\delta
=(\mathbf{P}_B\vec\epsilon)^\dag\vec\delta=0,
\eeq
using equation (\ref{eq:mpd=pd}), then the Hermiticity of $\mathbf{P}_B$,
then the fact that $\vec\epsilon$ is an E mode.
So $\mathbf{P}_B\vec\delta$ is indeed a pure B map.

For full-sky data, implemented in HEALPix for example \cite{healpix}, we can transform easily and quickly back and forth between the pixel basis and the spherical harmonic basis. The projections $\mathbf{P}_E,\mathbf{P}_B$ are trivial in the latter basis: they are diagonal matrices with ones and zeros along the diagonal.
Because of this, the problem above can be solved efficiently using conjugate gradient minimization.

%% We have already noted that this procedure would have worked just as well with  a different choice of inner product. In particular, we could have chosen to minimize $\phi = (\mathbf{P}_B\vec\delta)^\dag \mathbf{A}^{-1}
%% (\mathbf{P}_B\vec\delta)$ for any positive definite matrix 
%% $\mathbf{A}$. If $\mathbf{A}$ is diagonal in the spherical harmonic basis, then this corresponds to applying an $l$-space weighting when minimizing. As a result, there are in principle different ``pure'' maps for any given data set, corresponding to different choices of inner product. 

\begin{figure*}[t]
\centerline{\hfil\includegraphics[width=2.5in]{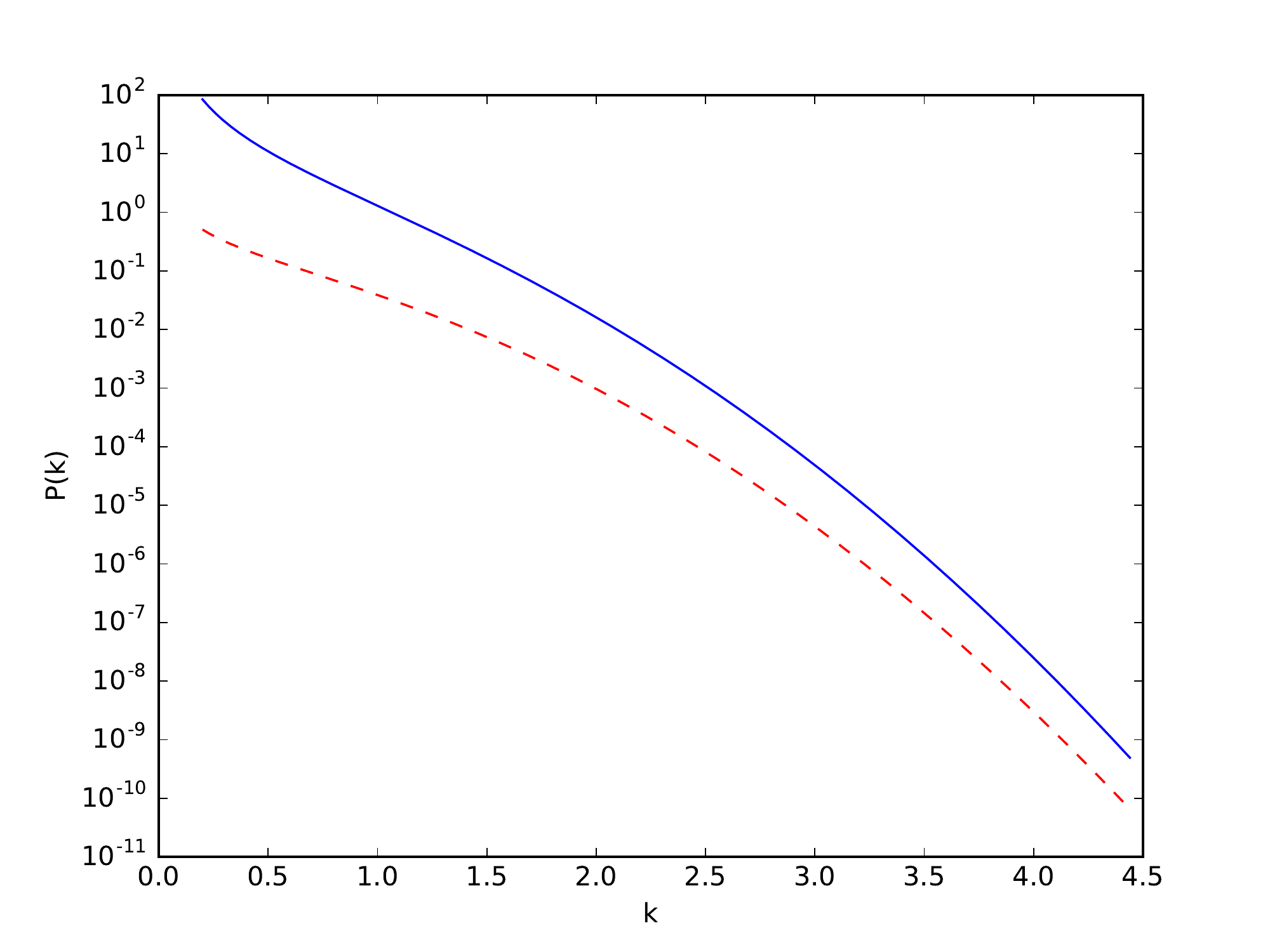}
\hfil
\includegraphics[width=2.5in]{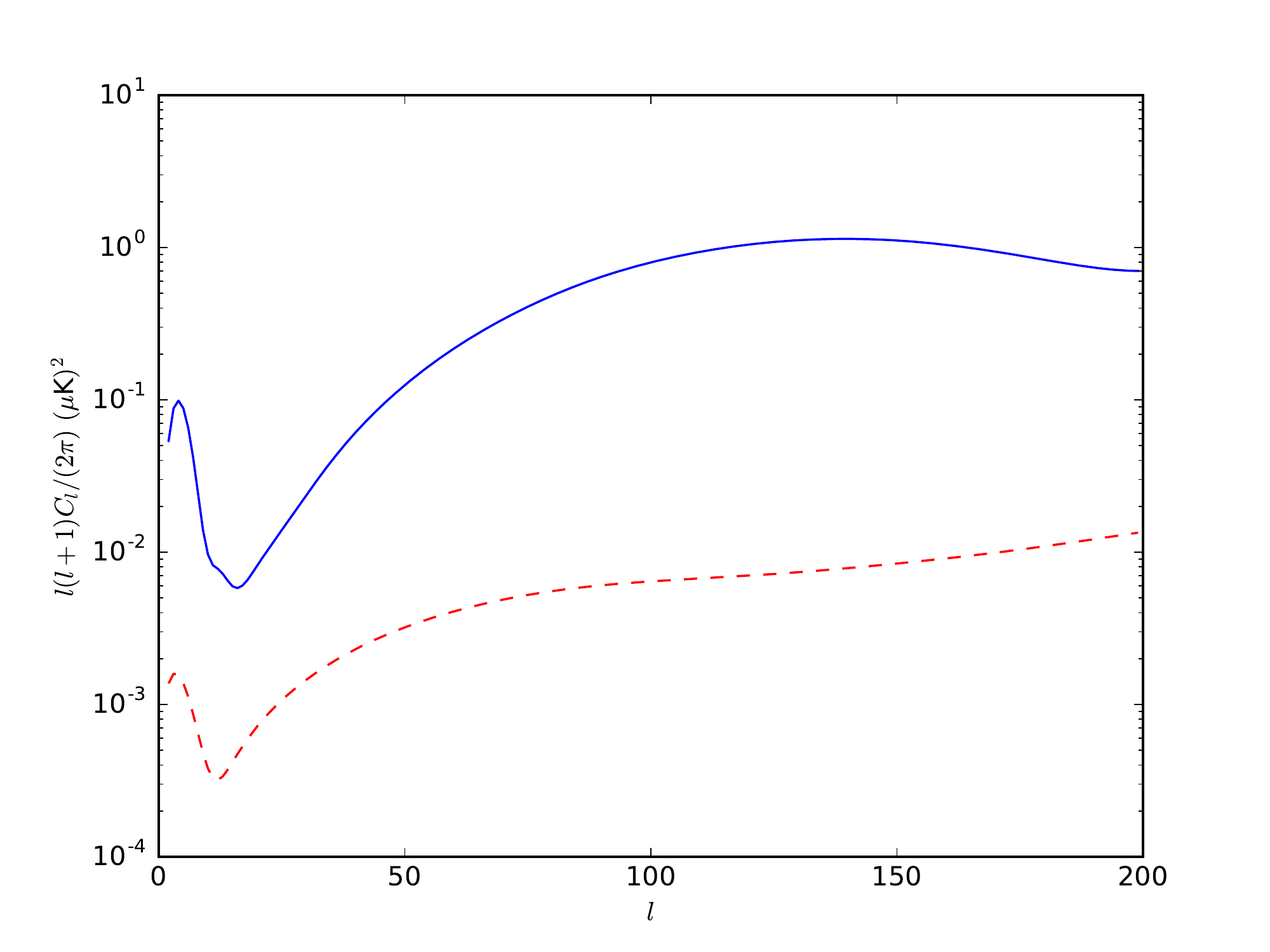}}
\caption{Power spectra used to create the data used in the examples described in Section \ref{sec:examples}. The left panel shows the power spectra used
in the flat-domain examples, and the right panel shows the spectra for the
spherical example. In both cases, the solid blue curve is the E spectrum, and the dashed red curve is the B spectrum.
}
\label{fig:pofk}
\end{figure*}

We can generalize this procedure by acknowledging the existence of noise in the maps. Instead of requiring strict agreement between $\vec\delta$ and $\vec d$ in the observed pixels, we can penalize disagreement in the usual way by 
introducing an inverse noise matrix $\mathbf{N}^{-1}$ 
and finding $\vec\delta_N$
that minimizes
\beq
\chi^2 = (\mathbf{P}_B\vec \delta_N)^2+
(\vec d-\vec \delta_N)\mathbf{N}^{-1}(\vec d-\vec \delta_N).
\eeq
The inverse noise matrix is diagonal, with 
$(\mathbf{N}^{-1})_{jj}=0$ for masked pixels. For observed
pixels, we set $\mathbf{N}^{-1}_{jj}=\sigma_j^{-2}$ for
some noise levels $\sigma_j$.
In the no-noise limit $\sigma_j\to 0$, discrepancies in the 
unmasked region are infinitely penalized, and 
the result 
corresponds to the pure map described above:
\beq
\vec\delta = \lim_{\sigma_j\to 0}\vec\delta_N.
\label{eq:nonoiselimit}
\eeq
One way to prove this formally is to write out $\partial\chi^2/\partial
\delta_{Nj}=0$ for both masked and unmasked pixels. For
unmasked pixels, the limit $\sigma_j\to 0$ enforces $\delta_{Nj}\to\delta_j$.
For masked pixels, it require $(\mathbf{P}_B\vec\delta)_j=0$ as required
by equation (\ref{eq:lagrangemultiplier}).

As we will see in the next section, this correspondence leads to a natural way of thinking of the pure E and B modes as applications of a Wiener filter.

%% Another generalization would be to minimize some other quadratic
%% function of $\mathbf{P}_B\vec\delta$,
%% \beq
%% \phi = (\mathbf{P}_B\vec\delta)^\dag\mathbf{A}(\mathbf{P}_B\vec\delta).
%% \eeq
%% By an argument like the one above, the vector $\mathbf{P}_B\mathbf{A}
%% \mathbf{P}_B\vec\delta$ is a pure B map (orthogonal to all E modes).

\section{Wiener filter}
\label{sec:wf}

We begin by recalling the general features of the Wiener filter. Assume that we have a data vector $\vec d=\vec s+\vec n$, where the signal $\vec s$ and noise $\vec n$ are Gaussian random vectors with zero mean and covariance matrices $\mathbf{S}$ and $\mathbf{N}$. The Wiener-filtered (WF) map is the signal $\vec s$ that maximizes the posterior probability. The probability density
of $\vec s$ is proportional to
$e^{-\chi^2/2}$, with
\beq
\chi^2 = \vec s^{\,\dag}\mathbf{S}^{-1}\vec s+(\vec d-\vec s)^\dag\mathbf{N}^{-1}
(\vec d-\vec s).
\eeq
The WF map is the vector $\vec s_W$ that minimizes this:
\beq
\vec s_W=(\mathbf{S}^{-1}+\mathbf{N}^{-1})^{-1}\mathbf{N}^{-1}\vec d
=\mathbf{S}(\mathbf{S}+\mathbf{N})^{-1}\vec d.
\label{eq:wforiginal}
\eeq

If $\mathbf{S}$ contains the covariances of both E and B signal, then the WF map contains the maximum-likelihood estimates for the combined E and B signal. One way to isolate one of these components is to treat the other as if it were a source of noise. For instance, we can get the WF B map by replacing $\mathbf{S}$ by $\mathbf{S}_B$
and $\mathbf{N}$ by $\mathbf{S}_E+\mathbf{N}$:
\begin{align}
\vec s_{W}^{\,B} &= (\mathbf{S}_B^{-1}+(\mathbf{S}_E+\mathbf{N})^{-1})^{-1}
(\mathbf{S}_E+\mathbf{N})^{-1}\vec d\\
&=\mathbf{S}_B(\mathbf{S}_B+(\mathbf{S}_E+\mathbf{N}))^{-1}\vec d.
\label{eq:sbw}
\end{align}
A similar expression applies to the E signal.
Note that $\vec s_{W}=\vec s_{W}^{\,E}+\vec s_{W}^{\,B}$.

These maps include the ambiguous modes. 
In particular, in the limit of low noise, $\vec s_W=\vec s_W^{\,E}+
\vec s_W^{\,B}$ approaches $\vec d$. In practice, since our theory typically assumes that there is much more E power than B power, ambiguous modes with high signal-to-noise are assigned mostly but not entirely to the E map.

The idea behind ``pure'' modes is to be absolutely sure that there is no
cross-contamination. That is, a pure B map is one whose power cannot possibly
have come from E modes. One way to produce a WF pure B map
is to let the E-mode power tend to infinity. Define
\beq
\mathbf{S}(\alpha)=\mathbf{S}_B+\alpha\mathbf{S}_E
\eeq
to be the signal matrix with E power inflated by a factor $\alpha$.
Replace $\mathbf{S}_B+\mathbf{S}_E$ with $\mathbf{S}(\alpha)$
in equation (\ref{eq:sbw})
to get
\begin{align}
\vec s_{W}^{\,B}(\alpha)&=\mathbf{S}_B(\mathbf{S}(\alpha)
+\mathbf{N})^{-1}\vec d\\
&=\mathbf{S}_B\mathbf{S}(\alpha)^{-1}
(\mathbf{S}(\alpha)^{-1}+\mathbf{N}^{-1})^{-1}\mathbf{N}^{-1}\vec d,
\end{align}
In the limit $\alpha\to\infty$, 
the only modes that survive are those that lie in the
null space of $\mathbf{S}_E$. These are the ``pure B'' modes.

The easy way to evaluate expressions like these is to extend the vector $\vec d$ to cover the entire sky, so that we can quickly convert back and forth between the pixel and spherical harmonic bases. We assign
infinite noise to the unobserved pixels, so $(\mathbf{N}^{-1})_{jj}=0$ for
such pixels.

In the spherical harmonic basis, the signal
matrices are diagonal:
\begin{align}
\mathbf{S}_E&=\mbox{diag}(C_2^E,\ldots C_{l_{\rm max}}^E,0\ldots,0)\\
\mathbf{S}_B&=\mbox{diag}(0,\ldots,C_2^B,\ldots C_{l_{\rm max}}^B)
\end{align}
where we have ordered the spherical harmonic basis to have all E modes first, and there are $2l+1$ copies of each $C_l^Z$.
Therefore,
\begin{align}
\mathbf{S}(\alpha)^{-1}&=\mbox{diag}((\alpha C_2^E)^{-1},\ldots
(\alpha C_{l_{\rm max}}^E)^{-1},\nonumber\\
&\qquad
(C_2^B)^{-1},\ldots (C_{l_{\rm max}}^B)^{-1}).
\end{align}
In the limit $\alpha\to\infty$, the E terms go to zero. The 
resulting matrix is the pseudo-inverse $\mathbf{S}_B^+$, which
is the inverse of $\mathbf{S}_B$ within the subspace of B modes and zero in the orthogonal subspace of E modes.

The WF pure B map is therefore
\beq
\vec s_{W}^{\,pB}\equiv
\lim_{\alpha\to\infty}\vec s_{W}^{\,B}(\alpha)
=
\mathbf{S}_B\mathbf{S}_B^{+}
(\mathbf{S}_B^{+}+\mathbf{N}^{-1})^{-1}\mathbf{N}^{-1}\vec d.
\label{eq:pureb}
\eeq
Note that $\mathbf{S}_B\mathbf{S}_B^+=\mathbf{P}_B$ is the operator
that projects onto the B subspace.
We can describe this procedure in the following way: to get the WF pure B map, we apply the filter (\ref{eq:wforiginal}), assuming infinite signal in the E sector, and then apply the projection $\mathbf{P}_B$ to the result.

\begin{figure*}[t]
\centerline{\includegraphics[width=2.2in]{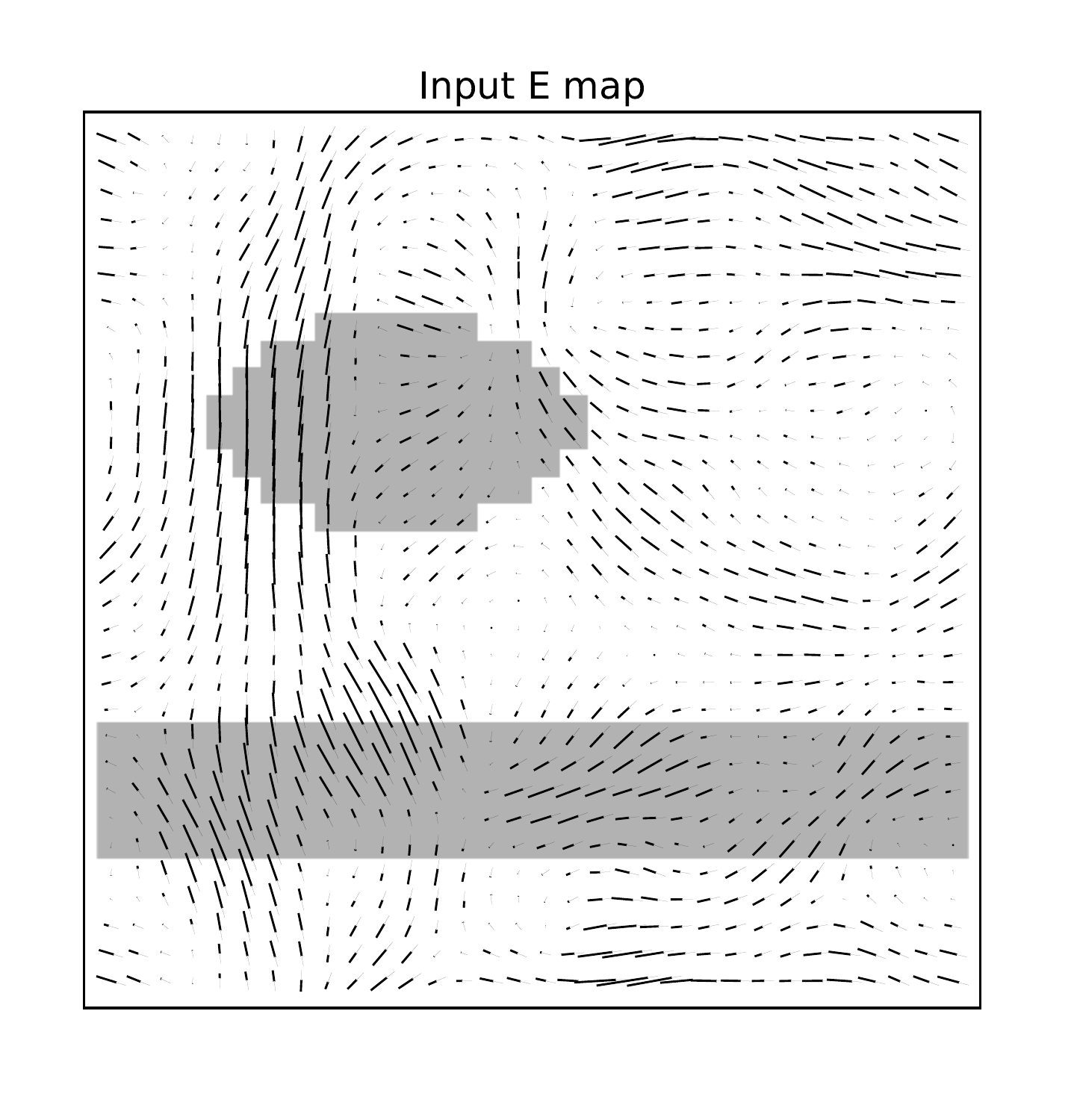}
\includegraphics[width=2.2in]{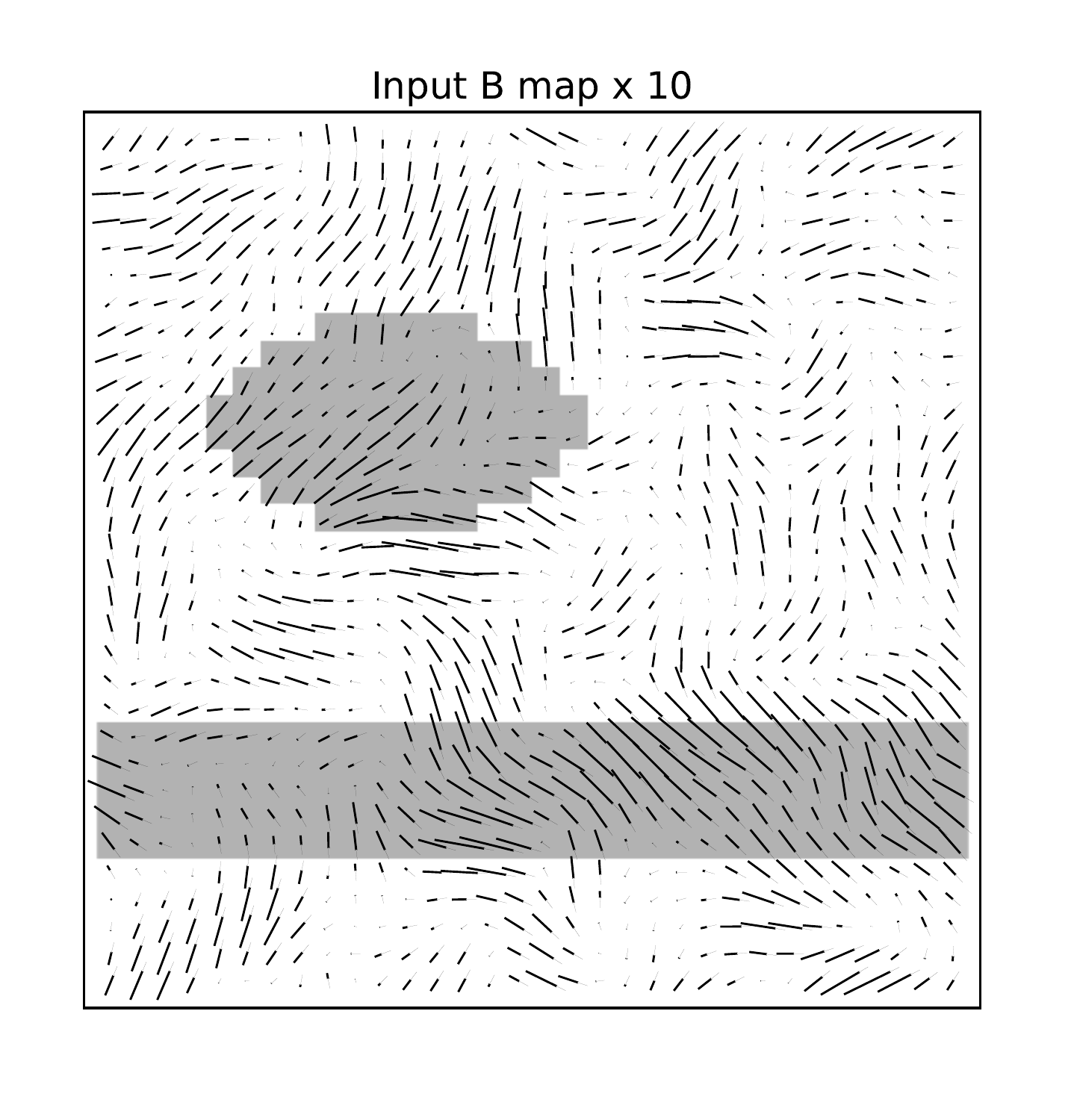}
\includegraphics[width=2.2in]{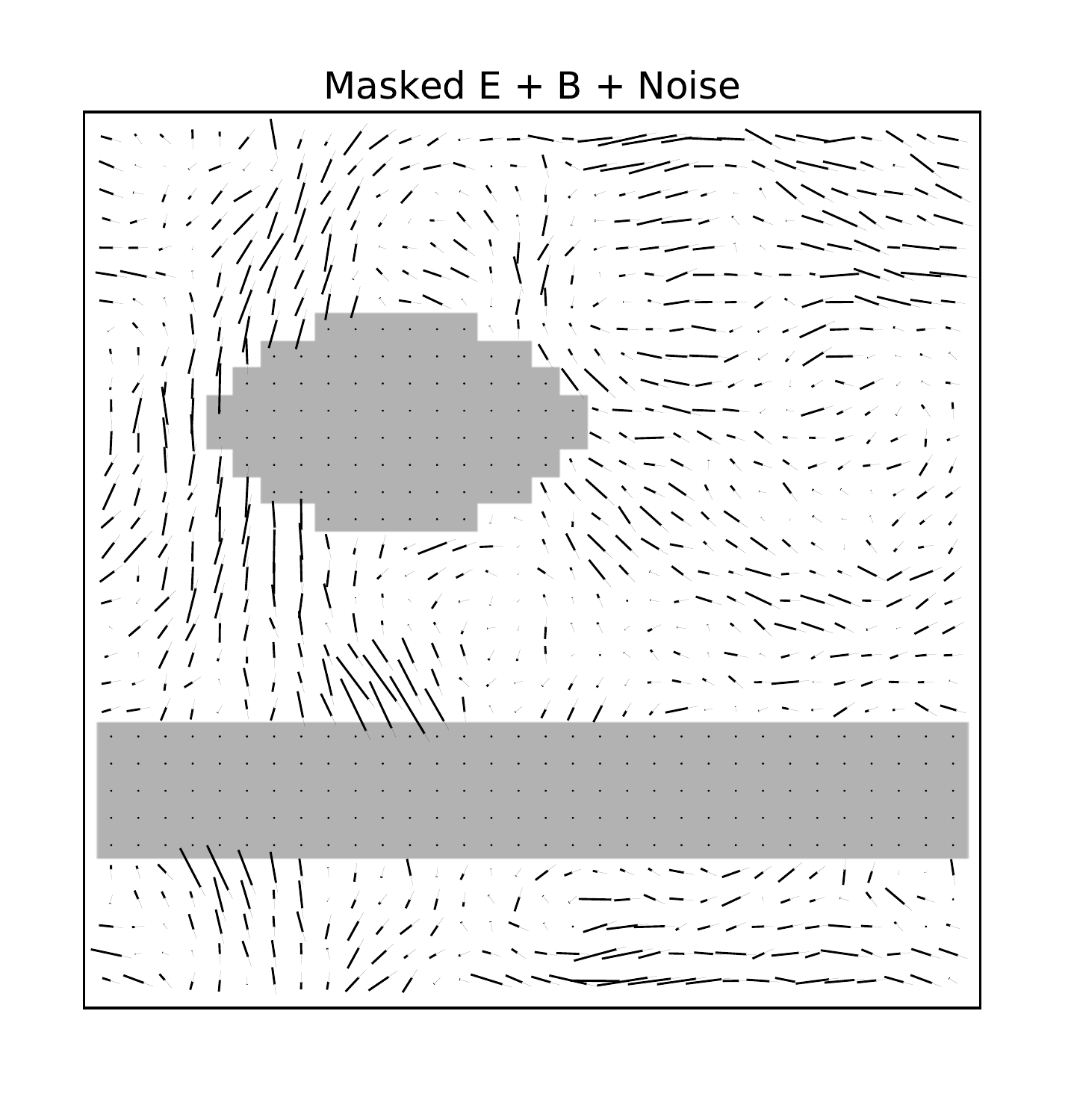}}
\caption{Example polarization maps: an E map (left), a B map (center), and a combined map containing the sum of the previous two, along with white noise. The noise amplitude is 0.3 times the signal. The gray region shows
the mask to be applied in subsequent filtering. Note that the B map
has been multiplied by 10 for visibility here and in subsequent figures.
}
\label{fig:inputs}
\end{figure*}

\begin{figure*}[t]
\centerline{
\includegraphics[width=2.2in]{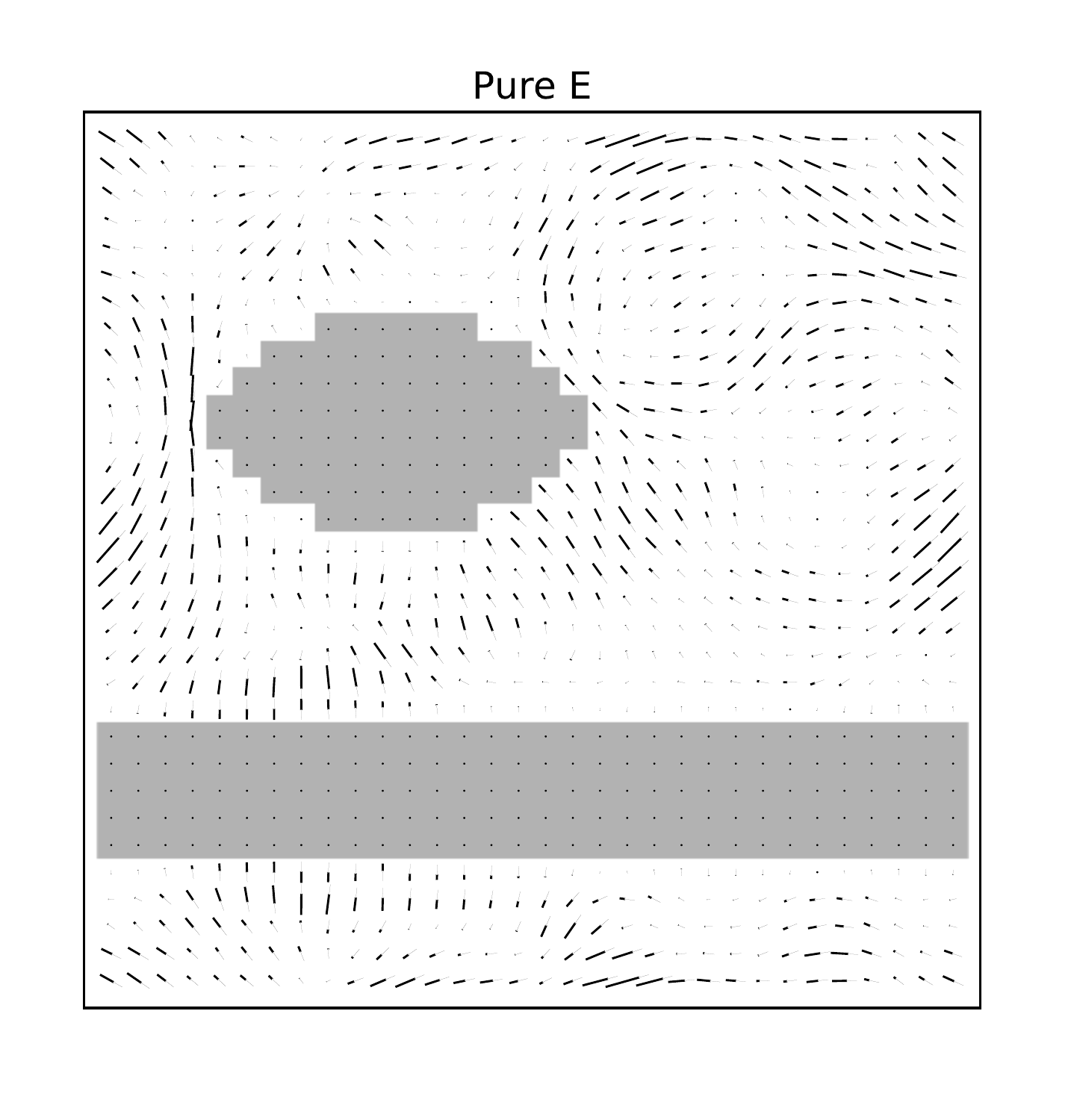}
\includegraphics[width=2.2in]{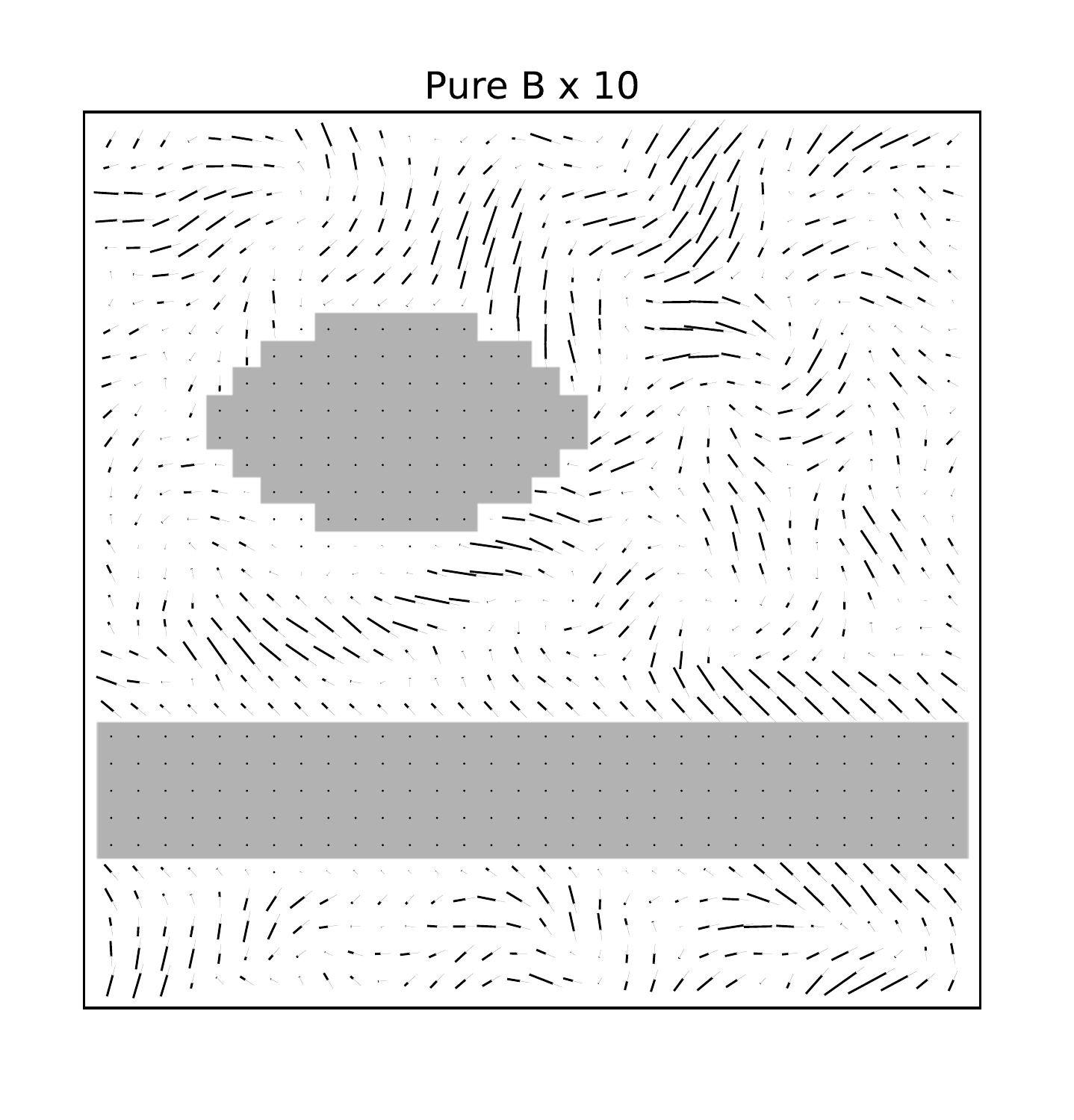}
\includegraphics[width=2.2in]{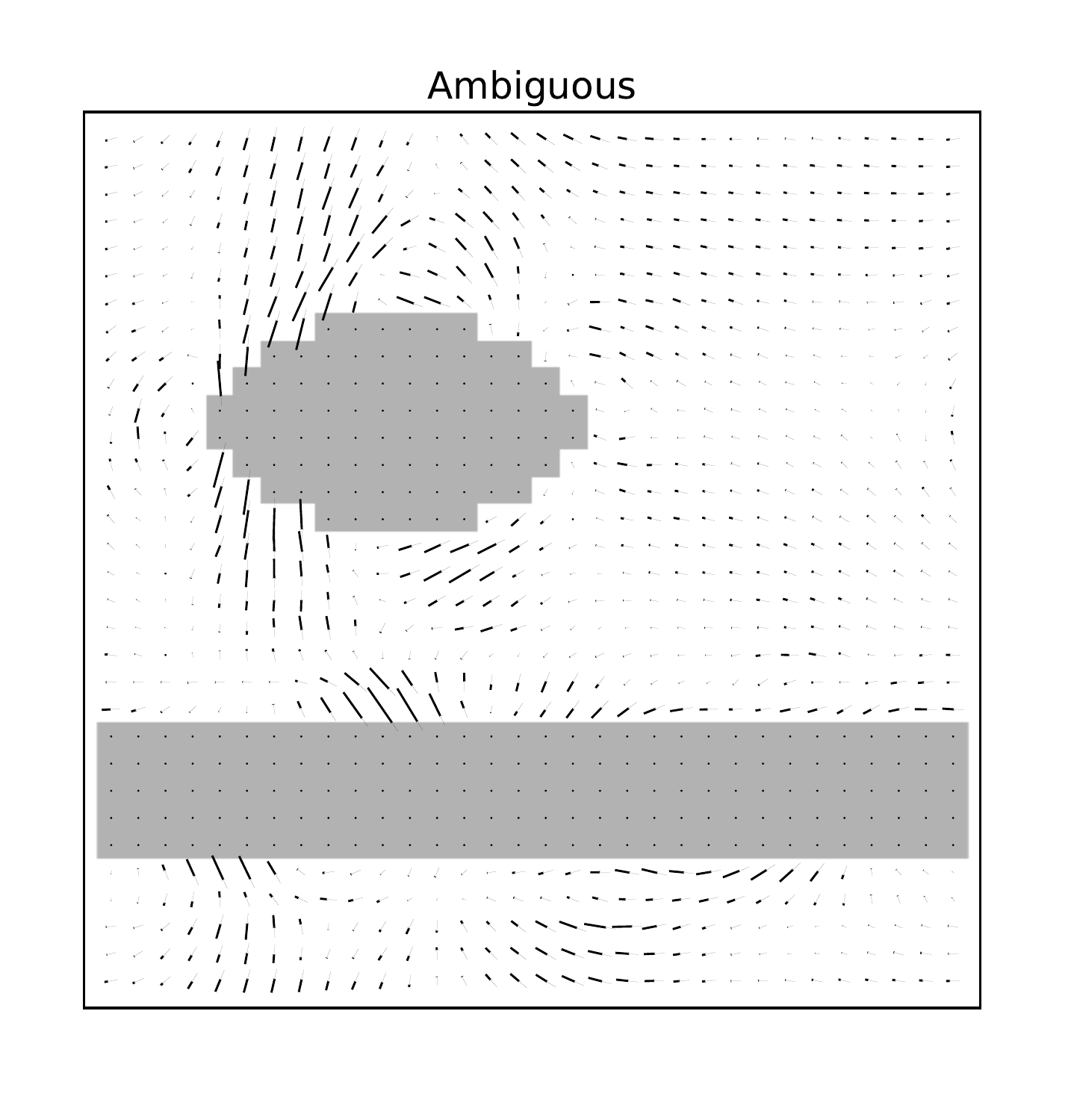}
}
\caption{
Decomposition of the map in Figure \ref{fig:inputs}. The input was the
combined E and B map, with no noise. The output pure E and B maps
were computed using the noise-free prescription in Section
\ref{sec:noisefree}. The ambiguous map is the difference between
the input map and the sum of the two pure maps.
}
\label{fig:noisefree}
\end{figure*}

The WF pure B map deserves the name ``pure,'' in the sense that it is derived entirely from the B modes of the true all-sky signal and is independent of any E modes. To see this, consider the unprojected filtered map
$(\mathbf{S}_B^++\mathbf{N}^{-1})\mathbf{N}^{-1}\vec d$. This map minimizes
$\chi^2=\vec s^{\,\dag}\mathbf{S}_B^+\vec s+(\vec d-\vec s)^\dag\mathbf{N}^{-1}(\vec d-\vec s)$. Now, suppose that $\vec d$ is derived from only E modes -- that is, it can be extended into the unobserved region in a way that has only E power. Then this extension will have $\chi^2=0$ and hence will be the map that minimizes $\chi^2$. When we then apply the projection operator
$\mathbf{P}_B$ onto the B subspace, the result will be zero. By linearity, therefore, any contribution to $\vec d$ that can be derived from E modes contributes zero to the WF pure B map.

Assuming uncorrelated noise, the matrix $\mathbf{N}^{-1}$ is diagonal in the pixel basis, while
$\mathbf{S}_B$ and $\mathbf{S}_B^+$ are diagonal in the spherical harmonic
basis, leading to efficient ways of evaluating this expression. In particular, the application of the matrix $(\mathbf{S}_B^++\mathbf{N}^{-1})^{-1}$ can be performed by either the conjugate gradient or the messenger method.

Suppose that we take the no-noise limit of the above procedure -- that is,
$(\mathbf{N}^{-1})_{jj}\to\infty$ for observed pixels and zero for unobserved
pixels. This forces perfect agremeent over the observed region, so it corresponds to a constrained minimization problem similar to the previous
section. Indeed, if we also adopt a white-noise (flat) power spectrum for $\mathbf{S}_B$, then the WF pure B map reduces precisely to the 
original pure B map. 

If we take the no-noise limit but do not adopt a white-noise power spectrum, we get different generalizations of the original pure B procedure, with different multipole-space weightings of the output. These maps are not ``pure'' by the original definition of ref.\ \cite{BZTO}: they are not orthogonal to all E modes with respect to the usual inner product. However, they do share the more important property of the original pure maps: they are sourced only by true B modes, with no contribution from E modes. 

\begin{figure*}[t]
\centerline{
\hfil
\includegraphics[width=2.2in]{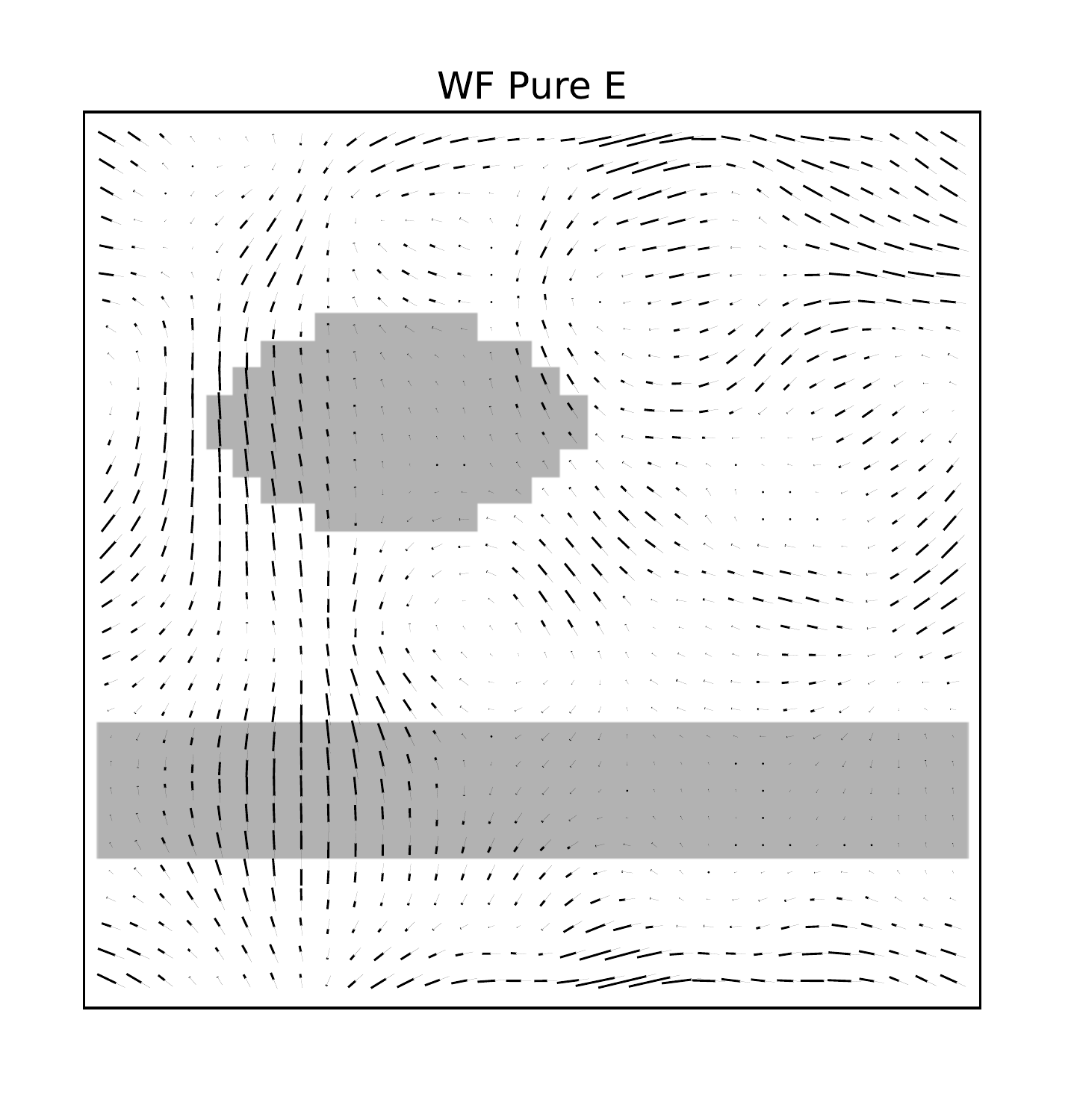}
\hfil
\includegraphics[width=2.2in]{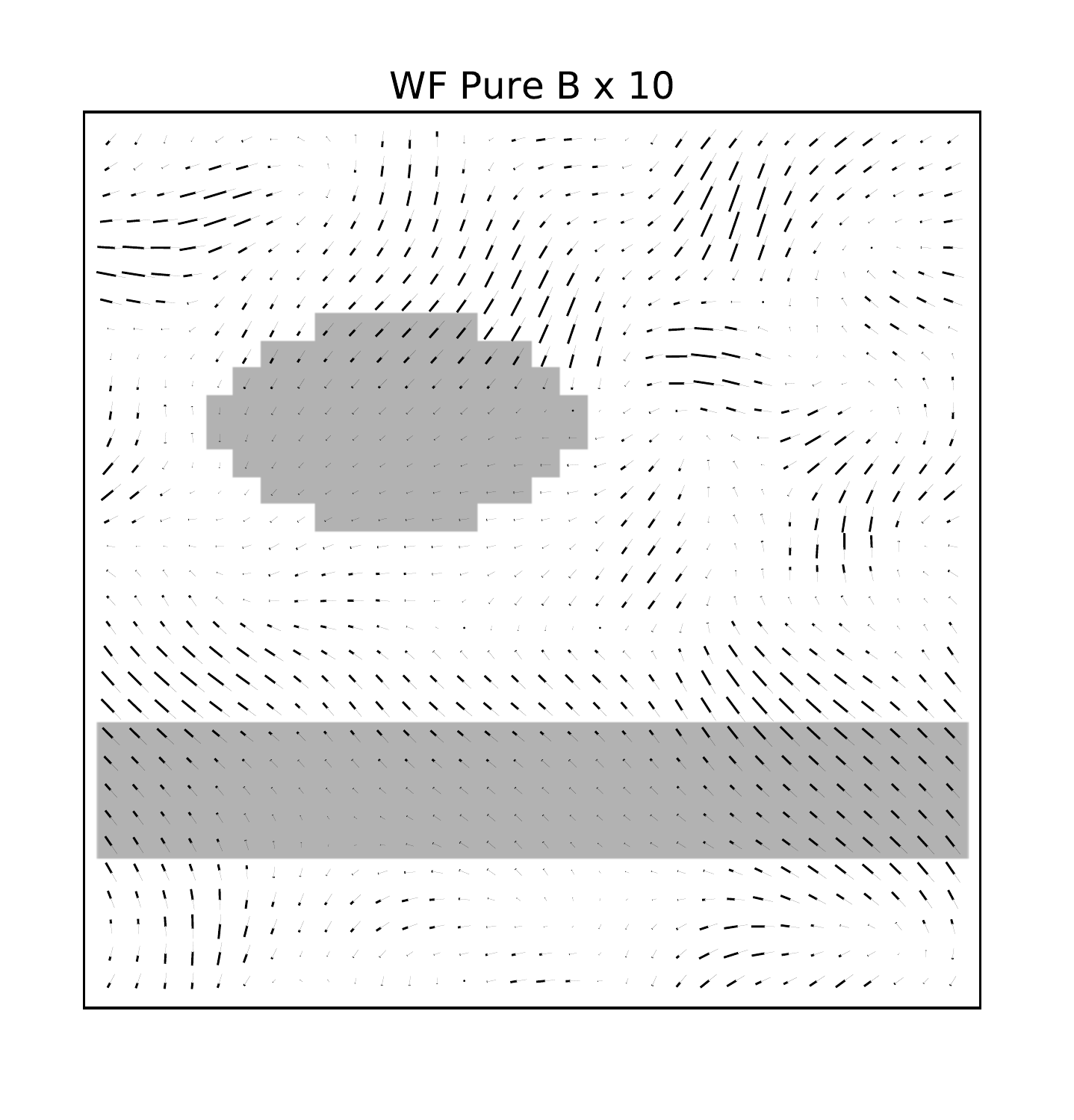}
\hfil
}
\caption{
Wiener-filtered pure E and B maps.
The input was the
combined E and B map, including noise (Fig. \ref{fig:inputs}). 
}
\label{fig:wf}
\end{figure*}

One useful application of the no-noise limit is to provide an initial guess for the full WF calculation.
The no-noise problem is lower-dimensional than the full WF (because observed pixels are fixed), so it can be solved quickly. For modes with high signal-to-noise, the result will be close to the WF map, so the output of this calculation can provide a good starting point for the conjugate gradient or messenger minimization.

\section{Correlation with temperature}
\label{sec:tcorr}

Thus far, we have assumed that the only relevant data are measurements of the linear polarization (Stokes $Q,U$). The polarization is correlated with the CMB temperature anisotropy (Stokes $I$). We now generalize the earlier results to account for this. 

Suppose that the data consists of both temperature and polarization measurements. We will write the data vector as $\vec d = \begin{pmatrix}
\vec d_T\\ \vec d_P\end{pmatrix}$, with all of the temperature measurements in $\vec d_T$ and all of the polarization measurements in $\vec d_P$. 

As
before, we will assume that these vectors cover the entire sky, assigning infinite
noise to unobserved pixels. Then the signal covariance matrix can
be written simply in the spherical harmonic basis.
If the spherical harmonic coefficients are in the order $T,E,B$,
then $\mathbf{S}$ can be written in block form as
\beq
\mathbf{S} = \begin{pmatrix} \hat{\mathbf{S}}_T & \hat{\mathbf{S}}_X & 0\\
\hat{\mathbf{S}}_X^\dag& \hat{\mathbf{S}}_E & 0\\
0 & 0 & \hat{\mathbf{S}}_B\end{pmatrix}.
\eeq
The matrix $\hat{\mathbf{S}}_T=\mbox{diag}(C_2^{TT},\ldots C_{l_{\rm max}}^{TT})$,
and $\hat{\mathbf{S}}_E,\hat{\mathbf{S}}_B,\hat{\mathbf{S}}_X$ similarly
contain the EE, BB, and TE power spectra. (The hats indicate that these are sub-blocks of the full covariance matrix. For instance, $\hat{\mathbf{S}}_B$ is the
nonzero block of the larger matrix $\mathbf{S}_B$ seen in the previous section.)

We can find the pure B map just as before, by replacing $\hat{\mathbf{S}}_E$
by $\alpha\hat{\mathbf{S}}_E$ and taking the limit $\alpha\to\infty$.
If we do this while leaving the other spectra, including the TE covariance,
fixed, then
it is straightforward to show that
\beq
\lim_{\alpha\to\infty}\mathbf{S}(\alpha)^{-1}=\begin{pmatrix}
\hat{\mathbf{S}}_T^{-1} & 0  & 0\\
0 & 0 & 0\\
0 &  0 & \hat{\mathbf{S}}_B^{-1}
\end{pmatrix}.
\eeq
The temperature and polarization sectors decouple, and the pure B
map is identical to the original expression (\ref{eq:pureb}).\footnote{
One might wonder whether it is correct to hold $\hat{\mathbf{S}}_X$
fixed while taking this limit. If we imagine that the ``extra'' E power associated with the inflated power spectrum came from a source uncorrelated
with T, then this is the correct procedure. If, on the other hand,
we take the limit while holding the TE correlation fixed, by
multiplying  $\hat{\mathbf{S}}_X$ by $\alpha^{1/2}$, then the upper
left block of $\mathbf{S}(\alpha)^{-1}$ changes to $\hat{(\mathbf{S}}_T
-\hat{\mathbf{S}}_X\hat{\mathbf{S}}_E^{-1}\hat{\mathbf{S}}_X^\dag)^{-1}$.
The TE and B sectors still decouple, and the pure B filter is the same.}

For the pure E map, the situation is somewhat different. If our goal
is to produce an E map that is free of B contamination, then we 
take the limit as the B power tends to infinity, and the resulting matrix is
\beq
\lim_{\alpha\to\infty}\mathbf{S}(\alpha)^{-1}=
\begin{pmatrix}
\begin{pmatrix}
\hat{\mathbf{S}}_T & \hat{\mathbf{S}}_X\\
\hat{\mathbf{S}}_X^\dag & \hat{\mathbf{S_E}}\end{pmatrix}^{-1} & 
\begin{matrix}0\\0\end{matrix}\\
\begin{matrix}0\ \ \  & 0\ \ \ \end{matrix} & 0\end{pmatrix}
\eeq
This says that the natural way to find the pure E map in the presence of temperature data is to apply the WF to the T and E signals jointly.
In the notation of the previous section, we find a combined ``pure TE map''
\beq
\vec s_W^{\,pTE}=
\mathbf{S}_{TE}\mathbf{S}_{TE}^+(\mathbf{S}_{TE}^++\mathbf{N}^{-1})
\mathbf{N}^{-1}\vec d,
\eeq
where $\mathbf{S}_{TE}$ is the signal covariance matrix containing both T and E signals and their covariance, and $\mathbf{S}_{TE}^+$ is the pseudo-inverse found by restricting to the subspace of T and E modes (i.e. all
B modes lie in the null space of both $\mathbf{S}_{TE}$ and $\mathbf{S}_{TE}^+$).

\begin{figure*}[t]
\centerline{
\hfil
\includegraphics[width=2.2in]{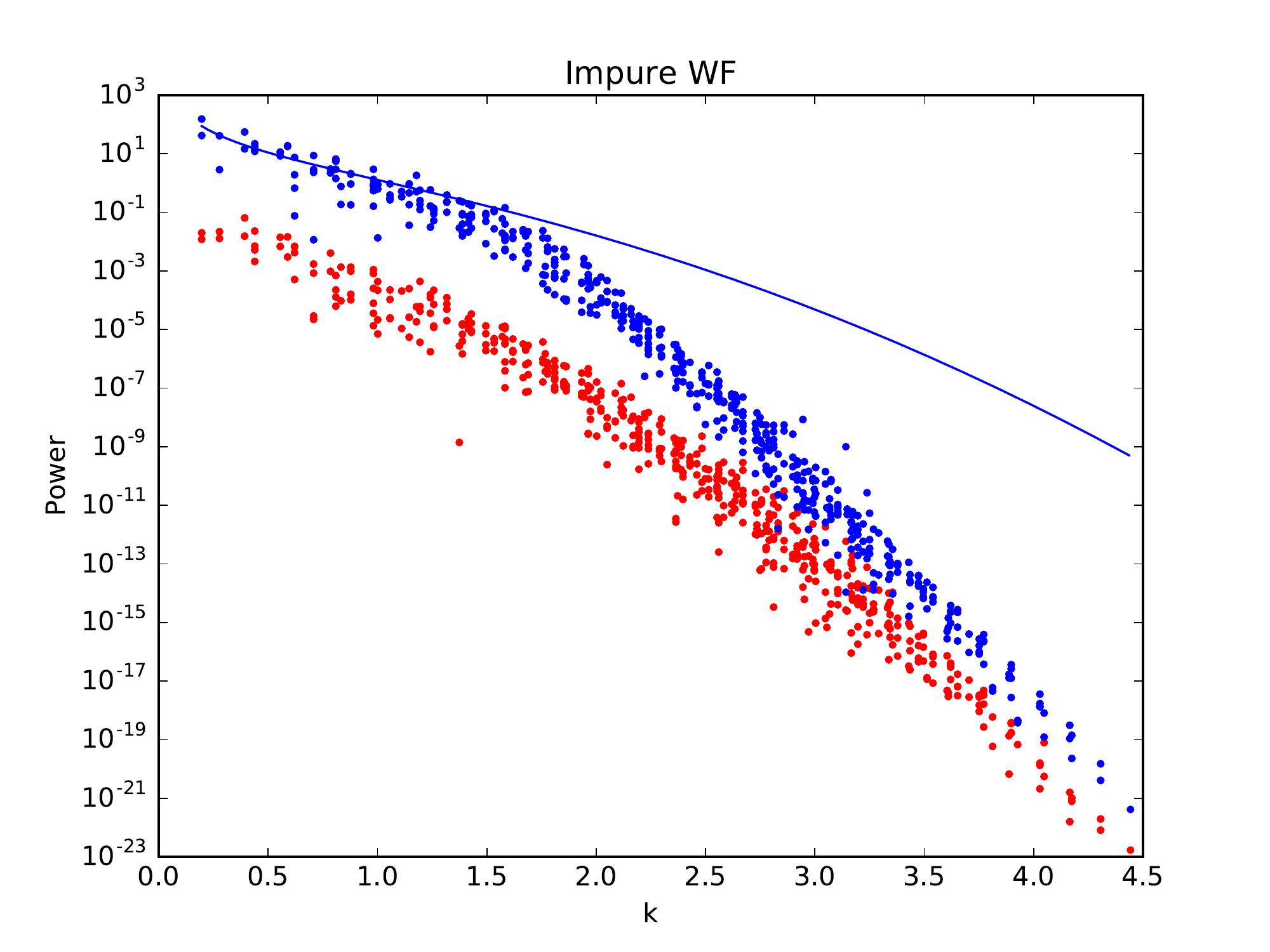}
\hfil
\includegraphics[width=2.2in]{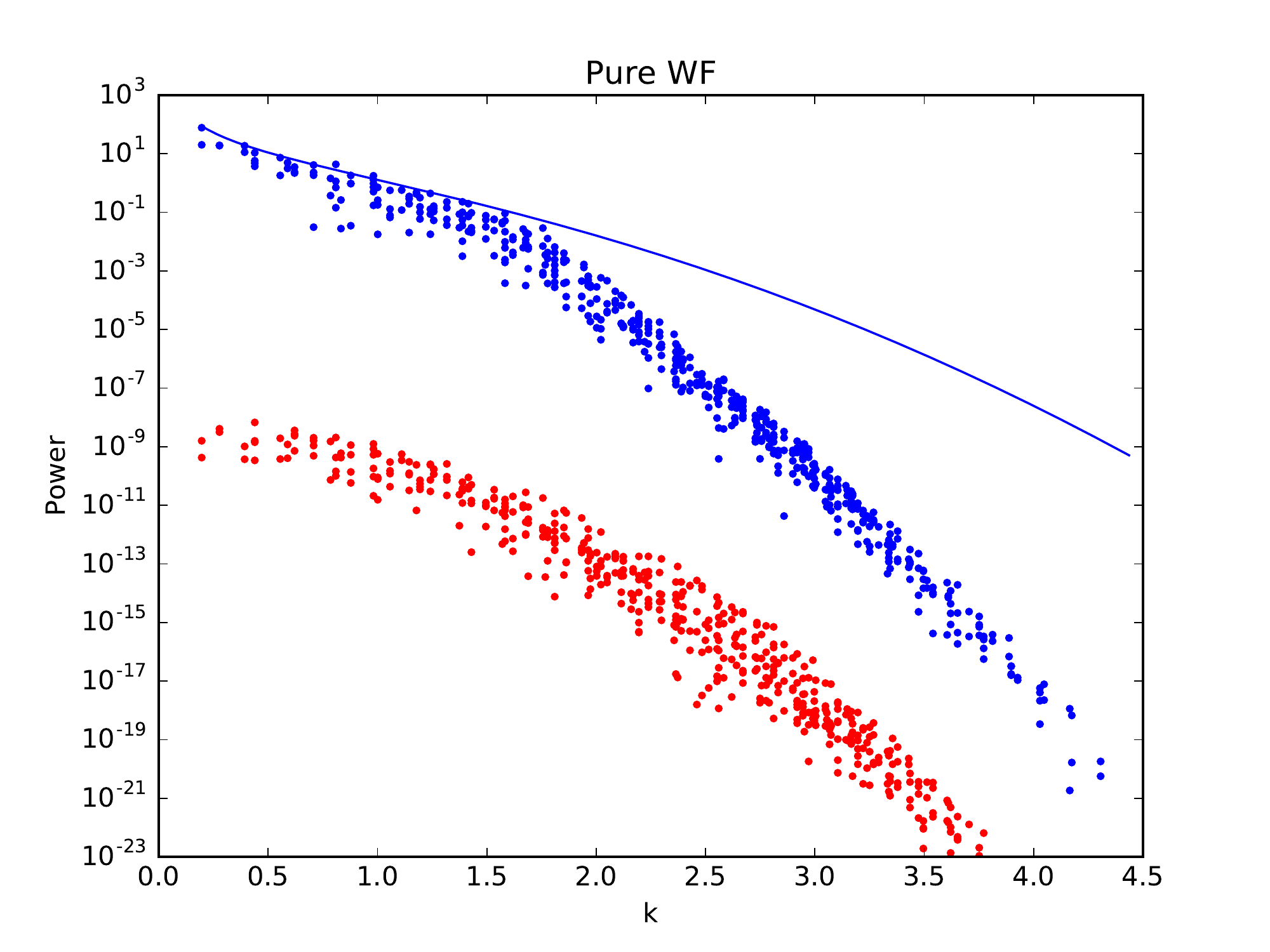}
\hfil
}
\caption{
The difference between the impure and pure Wiener filters.
The two panels show the results of applying the ``impure'' WF
(\ref{eq:wforiginal}) and the pure E and B WFs
(\ref{eq:pureb}) to an input map containing only the E polarization
signal shown in Figure \ref{fig:inputs}, with no B modes or noise.
The solid curve is the E power spectrum (see Fig. \ref{fig:pofk}). 
The points are the mean-square Fourier coefficients of the 
filtered maps.
In the left panel, the points are $|\tilde s_{W}^E(\vec k)|^2$ (blue)
and $|\tilde s_W^{B}(\vec k)|^2$ (red). The right panel shows
$|\tilde s_W^{pE}(\vec k)|^2$ and $|\tilde s_W^{pB}(\vec k)|^2$.
The nonzero signal in the pure B map is consistent with the numerical
accuracy of the conjugate-gradient minimization procedure.
}
\label{fig:impurevspure}
\end{figure*}

Of course, ``purifying'' the TE map from contamination by B modes is typically of limited interest. A more useful pure map E map would be one that had been purified with respect to T -- that is, a map that contains only the part of the E polarization that is not predicted by the temperature data.
The procedure here would be to take the T power to infinity, holding the TE correlation fixed. This corresponds to replacing
$\hat{\bf S}_T,\hat{\bf S}_X$ by $\alpha\hat{\bf S}_T,\alpha^{1/2}\mathbf{S}_X$
respectively. To get a pure E map, we should simultaneously purify with respect to B. 
In this case, 
\beq
\mathbf{S}_{TE}^+\equiv\lim_{\alpha\to\infty} \mathbf{S}(\alpha)^{-1}=
\begin{pmatrix}
0 & 0 & 0\\
0 & (\hat{\mathbf{S}}_E-\hat{\mathbf{S}}_X^\dag\hat{\mathbf{S}}_T^{-1}
\hat{\mathbf{S}}_X)^{-1} & 0 \\
0 & 0 & 0
\end{pmatrix}.
\eeq
The filtered map is
\beq
\vec s_W^{\,pE}=\mathbf{S}_E\mathbf{S}_{TE}^+
(\mathbf{S}_{TE}^++\mathbf{N}^{-1})^{-1}\mathbf{N}^{-1}\vec d.
\eeq

\section{Examples}
\label{sec:examples}

In this section we illustrate some of the procedures described above.
For simplicity, we begin by considering examples that lie in a flat square domain, using Fourier transforms
instead of spherical harmonic transforms.  We present an example on a spherical sky at the end of this section. 

For the flat examples, the ``full sky'' is a square with
periodic boundary conditions, pixelized into a $32\times 32$ grid. Because this domain has no boundary, there are no ambiguous modes. The EB decomposition can be performed mode by mode in the Fourier basis: E modes
have polarization direction parallel and perpendicular to the wavevector,
and B modes have polarization oriented at $45^\circ$ angles. 
All minimizations and solutions of linear systems were performed with  conjugate gradient methods.

We 
adopt the E and B power spectra shown in the left panel of Figure \ref{fig:pofk}. The
$E$ spectrum is $P_E(k)\propto k^{-2}e^{-k^2\sigma_b^2}$, and
the B spectrum is $P_B(k)\propto k^{-1}e^{-k^2\sigma_b^2}$. The exponential
terms correspond to smoothing with a Gaussian beam of width $\sigma_b=1$ pixel (an FWHM of about 2.4 pixels). (We write $P(k)$ rather than $C_l$
to emphasize that these calculations are on a flat patch rather
than the sphere.)
The power spectra are normalized so that the rms signal per pixel is 1, and the ratio of rms E to rms B is 10.

Figure \ref{fig:inputs} shows a realization of Gaussian random polarization maps made with these two spectra. The left and center panels show the E and B maps respectively. (Note that the B map has been multiplied by 10 for
visibility.) The right panel shows the sum of the two, with the addition of noise at a level of 0.3 times the rms signal. The gray region shows the mask -- i.e., the area that will be presumed to be unobserved in the subsequent analysis.

Figure \ref{fig:noisefree} shows the results of applying the noise-free purification procedure to these maps. As found in previous work, the ambiguous modes have support primarily close to the mask (except for a small number of large-scale modes), so the pure maps are suppressed near the mask.

For comparison, we calculated the pure E and B maps by computing an eigenbasis and projecting as described in \cite{BZTO}. The rms difference in the pure maps computed by the two methods differ by about 2\%.

We verified that the results of the WF procedure reduces to the original pure maps in the low-noise limit [equation (\ref{eq:nonoiselimit})] by 
applying the WF purification procedure to the same maps, assuming a flat
power spectrum and a low noise level $\sigma=10^{-3}$. The rms
differences between
the pure E and B maps resulting from this procedure and the 
maps shown in Figure \ref{fig:noisefree} were $6\times 10^{-7}$
(E) and $7\times 10^{-8}$ (B) times the rms input signal.

Figure \ref{fig:wf} shows the pure E and B WFs applied to the noisy, masked map of Figure \ref{fig:inputs}. The pure WF simultaneously suppresses modes with low signal-to-noise (i.e. high-frequency modes) and rejects ambiguous modes. 

\begin{figure*}[t]
\centerline{
\hfil
\includegraphics[width=1.7in]{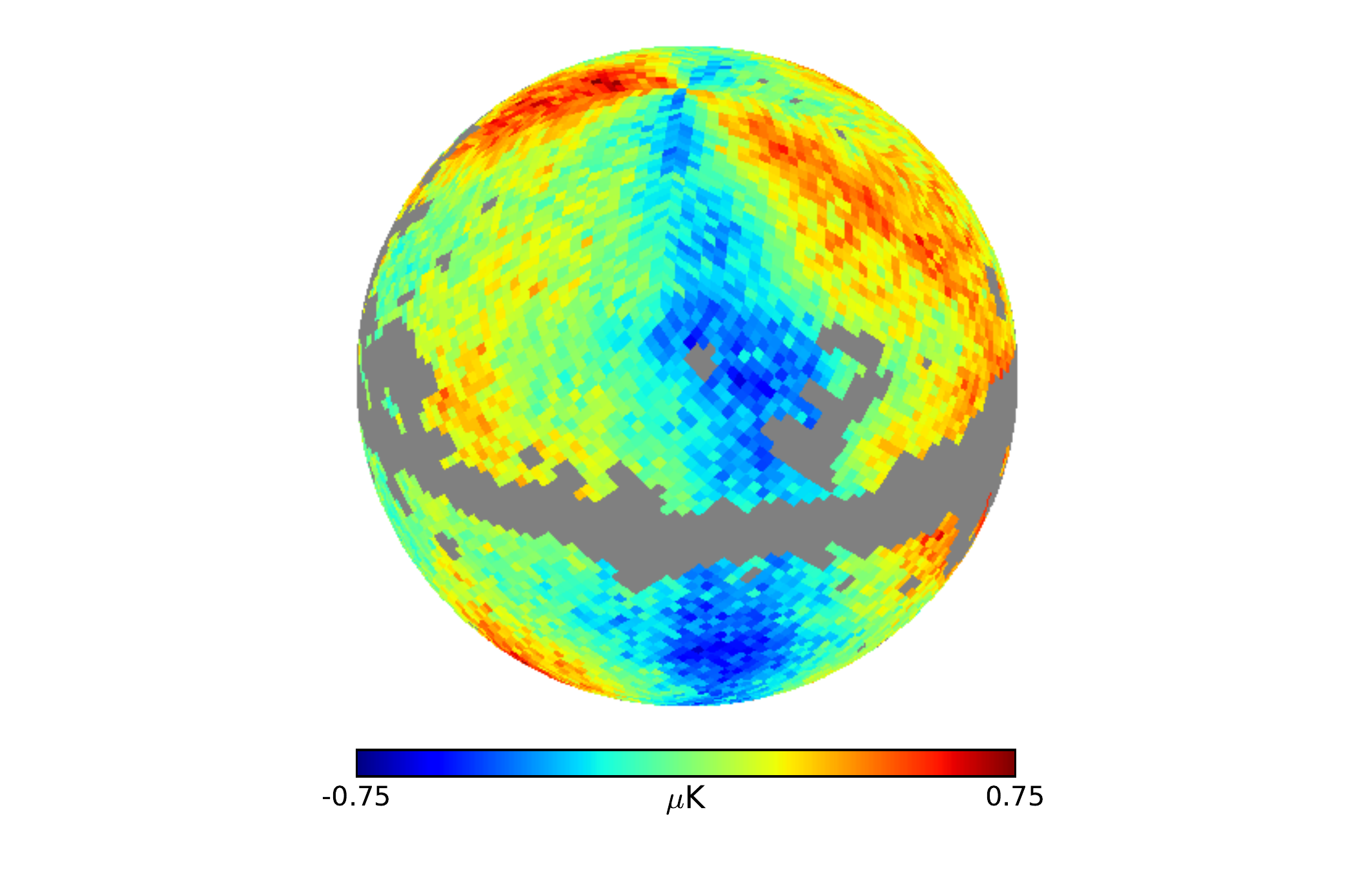}
\includegraphics[width=1.7in]{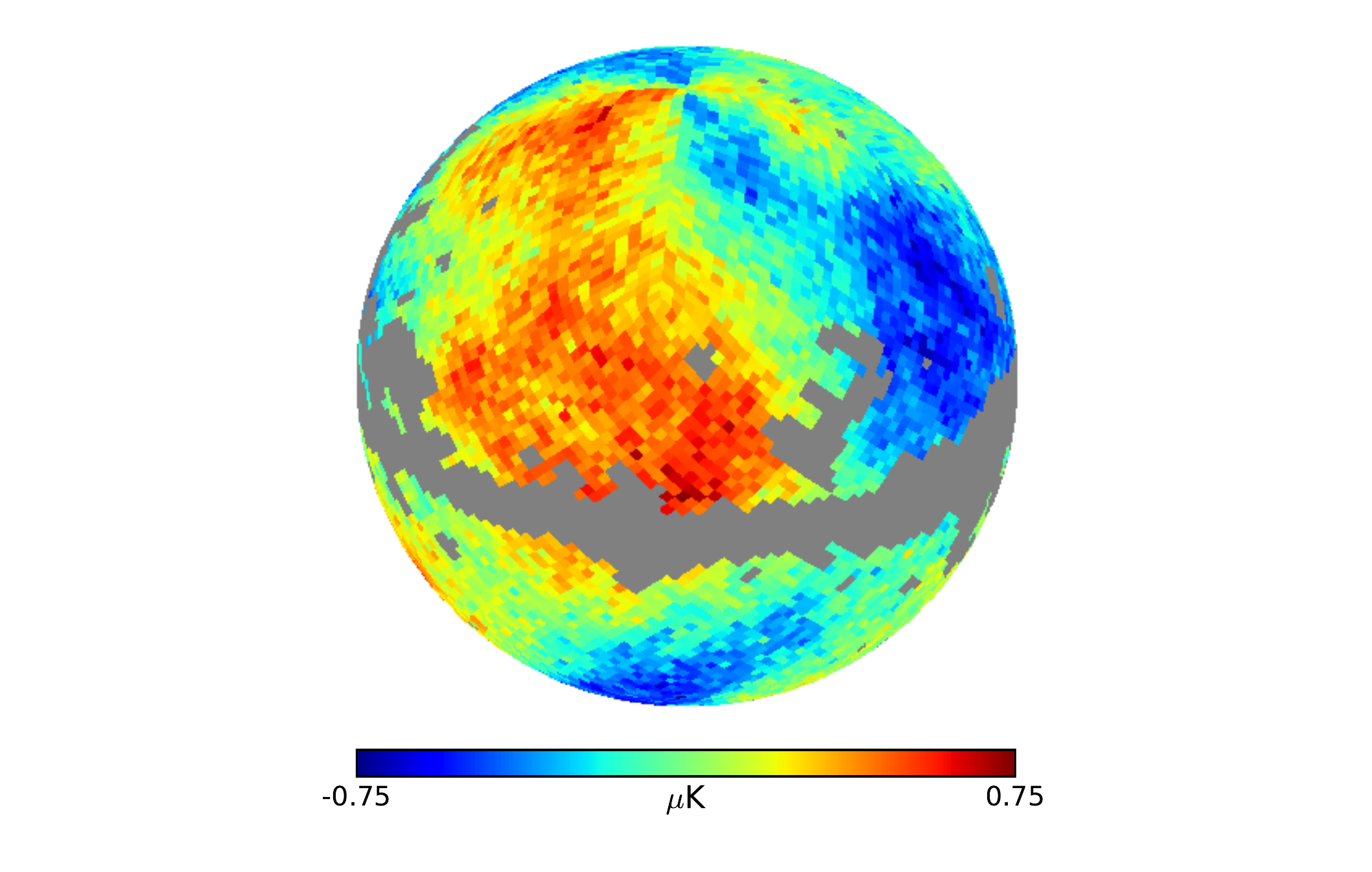}
\hfil
\includegraphics[width=1.7in]{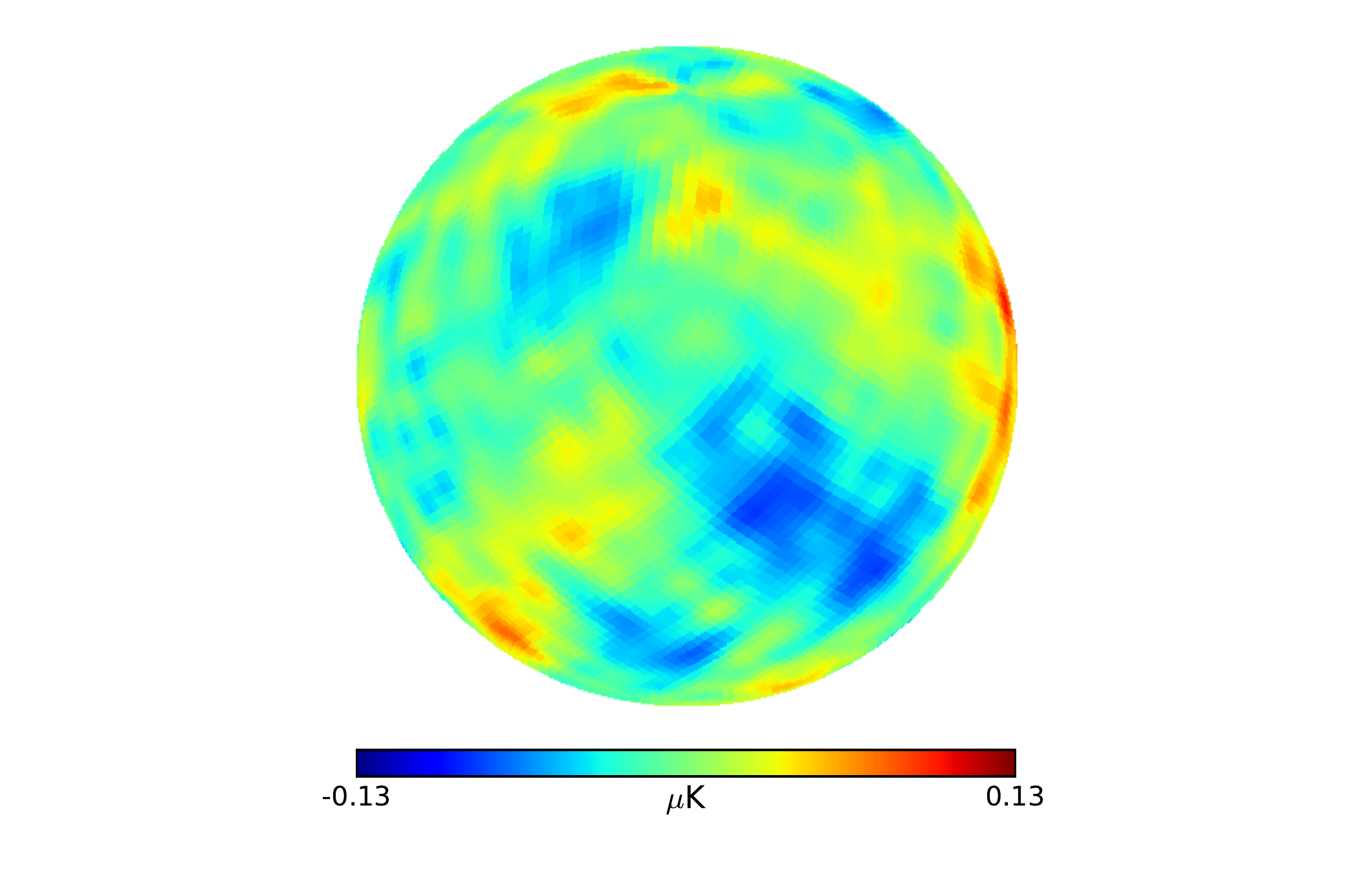}
\includegraphics[width=1.7in]{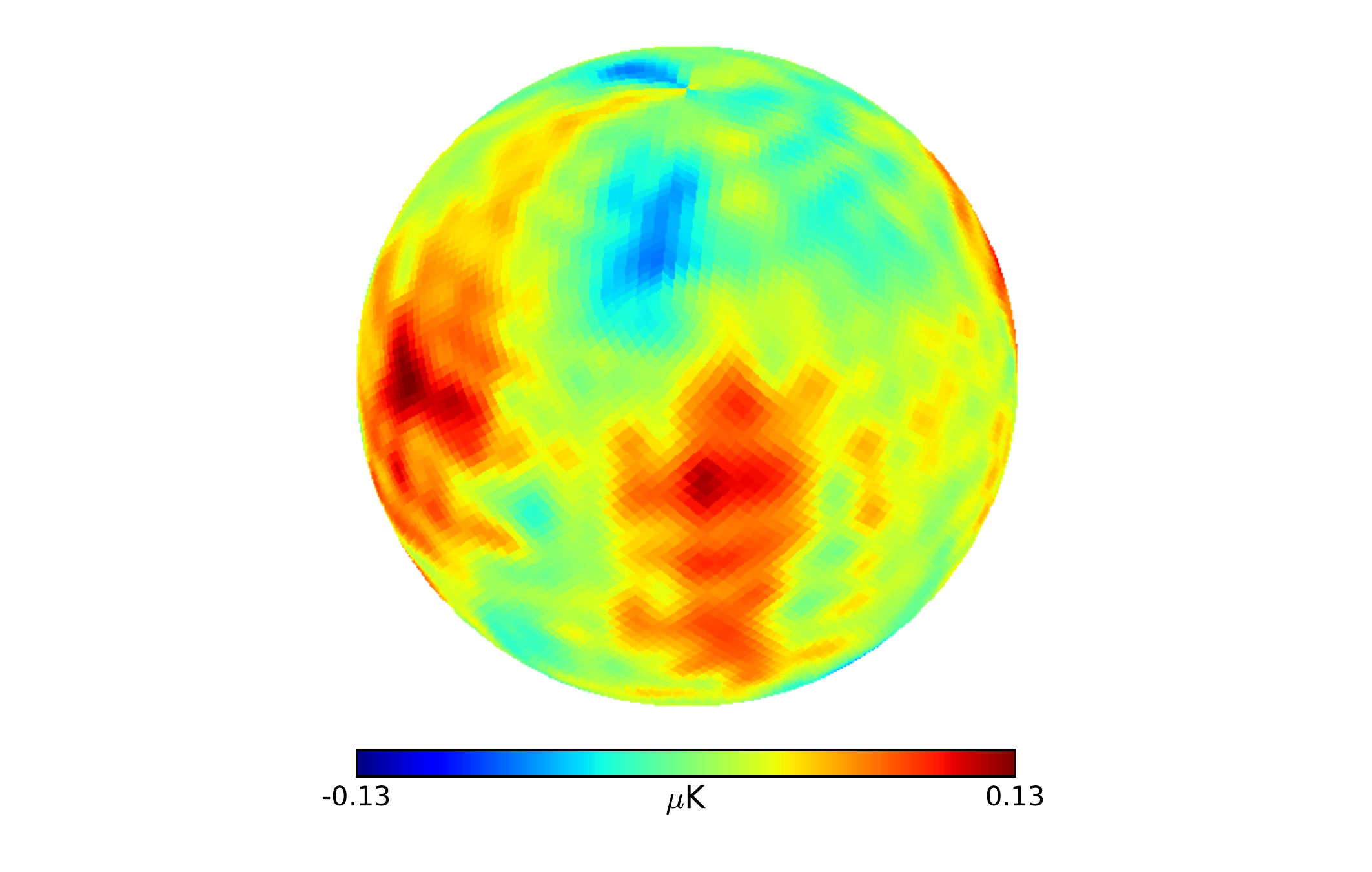}
}
\caption{Input data for tests of the pure WF on the sphere. All images
show orthographic projections of the southern equatorial hemisphere.
The left two panels are the masked $Q$ and $U$ maps, including E signal,
B signal, and noise. The right two panels show only the B component.}
\label{fig:sphere1}
\end{figure*}

Figure \ref{fig:impurevspure} illustrates the difference between the impure
and pure WFs.
In both panels, the input map is the E map from Figure \ref{fig:inputs}, with
no B modes or noise. The left 
panel shows the result of applying a WF without purification as in equation
(\ref{eq:wforiginal}). As expected, E modes with high signal-to noise (large scales) are almost fully recovered, while low signal-to-noise, small-scale modes are suppressed. The filtered B map is nonzero, showing E to B leakage.
The right panel shows the result of applying the pure E and pure B WFs.
The B power is negligible in this case.

We have also implemented the pure WF procedure on the sphere in the
HEALPix pixelization \cite{healpix}. We created a simulated polarization map based on the power spectra shown in the right panel of Figure \ref{fig:pofk}, which were calculated by CAMB \cite{camb} using
the best-fit Planck
cosmological parameters \cite{planckparams}, with a tensor-to-scalar ratio
$r=0.05$. A Gaussian beam with $\sigma_b=2.7^\circ$ was applied, and maps
of Stokes $Q$ and $U$ were made with HEALPix resolution $N_{\rm 
side}=32$. Gaussian noise was added with signal-to-noise of 5 in each pixel.
We assumed that only the Southern hemisphere had been observed, and additionally applied the Planck COMMANDER polarization mask \cite{planckmask},
resulting in a data set with 40\% sky coverage.
The resulting data set is shown in Figure \ref{fig:sphere1}. 

\begin{figure}[t]
\centerline{
\includegraphics[width=1.7in]{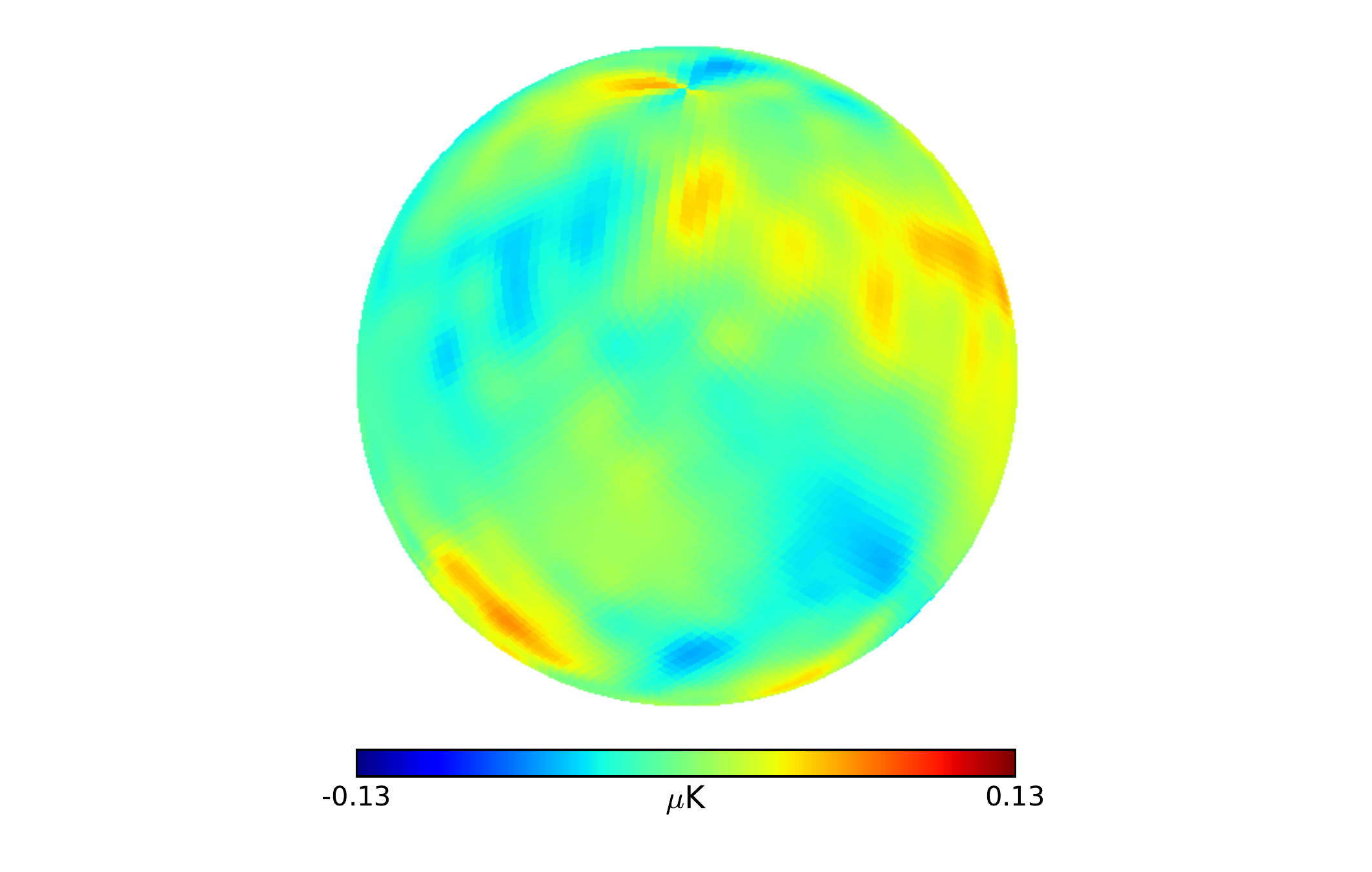}
\includegraphics[width=1.7in]{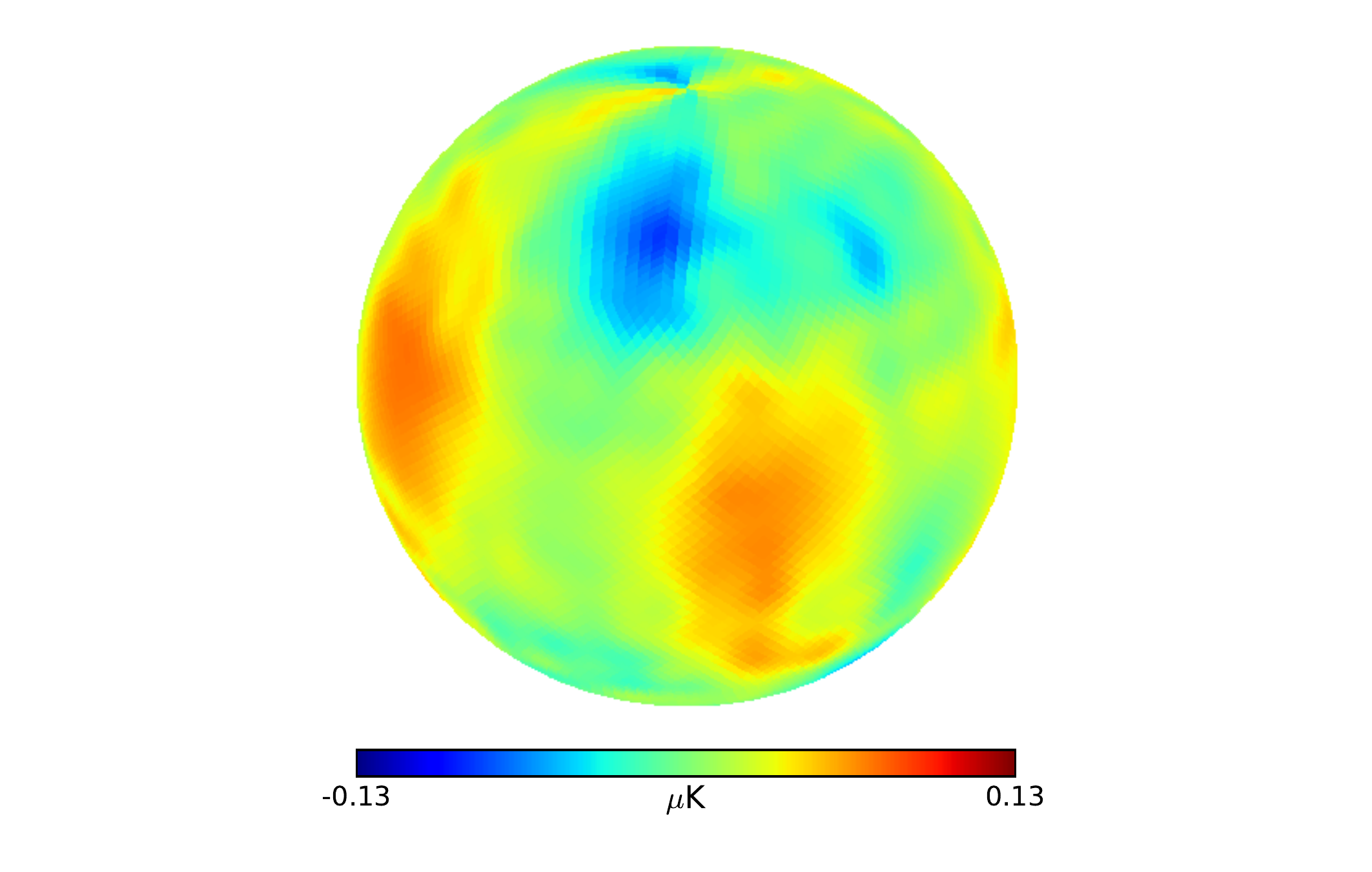}
}
\caption{Maps of Stokes $Q$ and $U$ for the pure B WF map derived from
the data shown in Figure \ref{fig:sphere1}.}
\label{fig:sphere2}
\end{figure}

The WF pure B maps are shown in Figure \ref{fig:sphere2}. 
As in the flat domain, the linear systems
are solved via preconditioned 
conjugate gradient methods, to take advantage of the
fact that the signal and noise matrices are diagonal in the harmonic
and pixel bases respectively. 
The preconditioner was taken to be diagonal in the harmonic
domain (as it would be for a map with all-sky coverage).
To speed convergence,
we set the E power
spectrum to 1000 times its true value (rather than to infinity), and we assumed
noise in the masked region of 20 times the rms signal (rather than infinity).
We verified that increasing these factors made negligible difference to
the final maps. 

We verified that the resulting map was pure by applying the WF to 
an input data set consisting of only B signal and noise, and to an
input containing only the E signal. As expected, the former led to a pure
B map virtually identical to the one shown, and the latter was mapped
nearly to zero, with rms fluctuations approximately 1\% of the pure B map.

\section{Conclusions}
\label{sec:conclusions}

Because the E and B components of a polarization map probe different
physics, it is important to be able to cleanly separate the
components. A cleaned B map that is guaranteed to be free of any
contamination from the larger E component is of particular value.  

The
methods we have presented provide efficient ways to calculate such
maps, based on the principle of the Wiener filter. 
The method generalizes the original notion of pure E and B maps, which considered only the effects of masking and pixelization, to account for noise simultaneously.
The method suppresses noise-dominated modes (as expected of a Wiener filter)
while guaranteeing a strict lack of contamination from E signal into the pure B map and vice versa.

The method can be implemented very efficiently, taking advantage of the sparsity of the E-B decomposition in the spherical harmonic basis. 

Because the method is based on maximizing the posterior probability, the filter is determined by the assumed power spectra, mask, and noise properties, with no arbitrary decisions such as apodization required. The resulting maps are filtered versions of the original polarization data, not scalar functions derived from them via a differential relation.

In addition to the original motiviation of eliminating E-B leakage, the method described herein can naturally be extended to include correlations with temperature data, producing, e.g., ``pure E'' maps that have had the temperature correlation removed.

The Wiener filter depends upon the choice of an input power spectrum. One can
choose a power spectrum estimate \textit{a priori}, or, as 
in methods
such as Gibbs sampling
\cite{LarsonGibbs}, samples of the power spectrum can be obtained based on the data itself. In this case, the probability distribution of pure WF maps and power spectra are obtained simultaneously. 

The ideas in this paper therefore generalize directly to power spectrum inference. Polarization auto- and cross-power spectrum inference as presented in \cite{LarsonGibbs} infers EE and BB spectra without purification of the maps:
the joint likelihood function $L(C_l^E,C_l^B)$ of the two spectra contains all relevant information, including E-B leakage. However, a detection of nonzero B power in the WF pure B map may be regarded as more robust than a detection derived from a joint analysis, as the WF pure B map ``tries as hard as it can'' to hide power coming from the E component.

The pure filtered maps are likely to be even more useful in contexts other than power spectrum estimation. Tests for foreground contamination, for example, or searches for non-Gaussianity and statistical anisotropy depend on real-space maps.
Perhaps most importantly, characterization of the B-type polarization produced via gravitational lensing depends on details of the B map that go beyond the power spectrum, and hence on real-space pure B maps.

\section*{Acknowledgments}

EFB is supported by NSF Award 1410133. This work has been done within the Labex ILP (reference ANR-10-LABX-63) part of the Idex SUPER, and received financial support managed by the Agence Nationale de la Recherche, as part of the programme Investissements d'avenir under the reference ANR-11-IDEX-0004-02.

\bibliography{purewiener}

\end{document}